\documentclass[notitlepage,nofootinbib,superscriptaddress]{revtex4}
\pdfoutput=1

\usepackage{amsmath,amssymb} 
\usepackage{graphicx}
\usepackage{xspace}	
\usepackage{siunitx}
\usepackage{booktabs}
\usepackage{xcolor}
\usepackage{tabularx}
\usepackage{multirow}
\usepackage{bm}
\usepackage{enumerate}
\usepackage{threeparttable}

\AtBeginDocument{
	\heavyrulewidth=.08em
	\lightrulewidth=.05em
	\cmidrulewidth=.03em
	\belowrulesep=.65ex
	\belowbottomsep=0pt
	\aboverulesep=.4ex
	\abovetopsep=0pt
	\cmidrulesep=\doublerulesep
	\cmidrulekern=.5em
	\defaultaddspace=.5em
}

\newcommand{\eg}{e.\,g.\xspace}
\newcommand{\ie}{i.\,e.\xspace}

\newcommand{\trapdistz}{\ensuremath{s_z}\xspace}
\newcommand{\trapdistx}{\ensuremath{s_x}\xspace}
\newcommand{\Wc}{\ensuremath{\mathit{\Omega}_\mathrm{c}}\xspace}

\newcommand{\Urf}{\ensuremath{U_\text{RF}}\xspace}
\newcommand{\Wrf}{\ensuremath{\mathit{\Omega}_\text{RF}}\xspace}
\newcommand{\wx}{\ensuremath{\omega_x}\xspace}
\newcommand{\wy}{\ensuremath{\omega_y}\xspace}
\newcommand{\wz}{\ensuremath{\omega_z}\xspace}
\newcommand{\wrad}{\ensuremath{\omega_\text{r}}\xspace}
\newcommand{\wrado}{\ensuremath{\omega_\text{r,1}}\xspace}
\newcommand{\wradt}{\ensuremath{\omega_\text{r,2}}\xspace}
\newcommand{\Ub}{\ensuremath{U_\text{b}}\xspace}
\newcommand{\lDC}{\ensuremath{l_\text{DC}}\xspace}

\newcommand{\Umw}{\ensuremath{U_\text{mw}}\xspace}
\newcommand{\Umwl}{\ensuremath{U_\text{mw}^{(l)}}\xspace}
\newcommand{\Umwr}{\ensuremath{U_\text{mw}^{(r)}}\xspace}
\newcommand{\thetar}{\ensuremath{\theta_\text{r}}\xspace}
\newcommand{\thetaz}{\ensuremath{\theta_z}\xspace}

\newcommand{\Ca}{\ensuremath{^{40}\text{Ca}^{+}}\xspace}
\newcommand{\Gh}{\ensuremath{\mathit{\Gamma}_\text{h}}\xspace}

\usepackage[colorlinks=True,linkcolor=blue,citecolor=blue]{hyperref}

\begin{document}

\title{Two-dimensional linear trap array for quantum information processing}

\author{P.\,C. Holz}
\altaffiliation{Corresponding author}
\email{philip.holz@uibk.ac.at}
\affiliation{Institut f\"ur Experimentalphysik, Universit\"at Innsbruck, Technikerstra\ss e 25, A-6020 Innsbruck, Austria}

\author{S. Auchter}
\affiliation{Institut f\"ur Experimentalphysik, Universit\"at Innsbruck, Technikerstra\ss e 25, A-6020 Innsbruck, Austria}
\affiliation{Infineon Technologies Austria AG, Siemensstra\ss e 2, A-9500 Villach, Austria}

\author{G. Stocker}
\affiliation{Institut f\"ur Experimentalphysik, Universit\"at Innsbruck, Technikerstra\ss e 25, A-6020 Innsbruck, Austria}
\affiliation{Infineon Technologies Austria AG, Siemensstra\ss e 2, A-9500 Villach, Austria}

\author{M. Valentini}
\affiliation{Institut f\"ur Experimentalphysik, Universit\"at Innsbruck, Technikerstra\ss e 25, A-6020 Innsbruck, Austria}

\author{K. Lakhmanskiy}
\affiliation{Institut f\"ur Experimentalphysik, Universit\"at Innsbruck, Technikerstra\ss e 25, A-6020 Innsbruck, Austria}

\author{C. R\"ossler}
\affiliation{Infineon Technologies Austria AG, Siemensstra\ss e 2, A-9500 Villach, Austria}

\author{P. Stampfer}
\affiliation{KAI Kompetenzzentrum Automobil- und Industrieelektronik GmbH, Technologiepark Villach, Europastra\ss e 8, A-9524 Villach, Austria}

\author{S. Sgouridis}
\affiliation{Infineon Technologies Austria AG, Siemensstra\ss e 2, A-9500 Villach, Austria}

\author{E. Aschauer}
\affiliation{Infineon Technologies Austria AG, Siemensstra\ss e 2, A-9500 Villach, Austria}

\author{Y. Colombe}
\affiliation{Institut f\"ur Experimentalphysik, Universit\"at Innsbruck, Technikerstra\ss e 25, A-6020 Innsbruck, Austria}

\author{R. Blatt}
\affiliation{Institut f\"ur Experimentalphysik, Universit\"at Innsbruck, Technikerstra\ss e 25, A-6020 Innsbruck, Austria}
\affiliation{Institut f\"ur Quantenoptik und Quanteninformation, \"Osterreichische Akademie der Wissenschaften, Technikerstra\ss e 21\,A, A-6020 Innsbruck, Austria}

\date{\today}

\begin{abstract}
	
	We present an ion-lattice quantum processor based on a two-dimensional arrangement of linear surface traps. Our design features a tunable coupling between ions in adjacent lattice sites and a configurable ion-lattice connectivity, allowing one, \eg, to realize rectangular and triangular lattices with the same trap chip. We present detailed trap simulations of a simplest-instance ion array with $2\times9$ trapping sites and report on the fabrication of a prototype device in an industrial facility. The design and the employed fabrication processes are scalable to larger array sizes. We demonstrate trapping of ions in rectangular and triangular lattices and demonstrate transport of a $2\times2$ ion-lattice over one lattice period.
	
\end{abstract}

\keywords{quantum information processing, ion traps, surface trap array, ion lattice}

\maketitle

\section{Introduction}
\label{sec:intro}

Trapped ions are one of the most successful platforms for quantum information processing to date, with high gate fidelities and long coherence times \cite{Olm2007,Sch2013,Har2014,Gae2016,Wan2017}. Trapped-ion quantum processors have been used to implement quantum algorithms, such as Grover's and Shor's algorithms \cite{Fig2017,Mon2016}, to implement quantum error correction protocols \cite{Chi2004,Nig2014}, and have also been successfully employed as quantum simulators, for instance for the observation of many-body dynamical phase transitions \cite{Zha2017,Jur2017}, the simulation of particle-antiparticle generation in lattice gauge theories \cite{Mar2016} or to calculate molecular ground state energies of simple molecules \cite{Hem2018}. Currently, the biggest challenge for trapped-ion quantum processors is to scale-up the number of qubits.
One approach towards scalable systems is to distribute the quantum register over several trapping zones using a so-called QCCD-architecture (quantum charge-coupled device) \cite{Win1998,Kie2002,Haf2008} and to eventually modularize the processor \cite{Bro2016,Lek2017,Bru2019}. In such a QCCD processor, each trapping zone contains only a small number of ions that can be manipulated with high-fidelity, while exchange of quantum information among the zones requires splitting, shuttling and merging of ion strings \cite{Win1998,Kie2002,Haf2008,Pin2020}. Complementary to this approach, one can also couple and entangle ions in different trapping potentials utilizing adiabatic well-to-well interactions \cite{Bro2011,Har2011,Wil2014,Hak2019} or, as recently proposed, broadband pulse sequences with high-power lasers \cite{Rat2018}. Following this idea, microfabricated ion trap arrays have been realized in order to create two-dimensional ion lattices on a microchip \cite{Cla2009,Ste2014,Tan2014,Bru2016,Kum2016,Mie2016}, and recently first quantum simulations have been performed in such a system \cite{Kie2019}. Arrays of individual traps have the advantage that the ions are not subjected to excess micromotion, in contrast to two-dimensional ion lattices naturally forming in a single trapping potential \cite{Ric2016,ivo2020}. In addition, microfabricated trap arrays offer a much finer control of the confining potential landscape and allow one to set the structure of the ion lattice by choice of the electrode geometry and control voltages.

Previous designs of ion-lattice quantum processors have mainly investigated surface \emph{point} traps as fundamental building block of the trap array \cite{Cla2009,Ste2014,Bru2016,Kum2016,Mie2016}. An exception is the work of Tanaka et al. \cite{Tan2014}, where two parallel \emph{linear} traps have been used. In our article, we further develop this latter approach: realizing a scalable trapped-ion quantum processor based on parallel linear surface traps. Our design enables tunable site-to-site coupling by combining the concepts of variable radio-frequency (RF) voltages \cite{Kum2011,Tan2014} with island-like electrodes, where static (DC) voltages are applied. The usage of linear traps as building block of the array thereby offers additional advantages compared to point traps. First, ions can be shuttled along the linear trap axes, giving rise to a configurable ion lattice connectivity and allowing for transport of quantum information, physically encoded in the ions, over large distances through the lattice. These points will be further discussed in the next sections. Second, due to the linear nature of the RF traps, multiple ions can be trapped in each site of the ion lattice without subjecting them to excess micromotion. Storage of multiple ions per site could be useful to increase the dipole-dipole coupling across adjacent sites \cite{Har2011}, thereby reducing the gate time for inter-site entangling operations. Furthermore, one could use sympathetic cooling techniques and apply standard quantum gate operations to ions within one site \cite{Hom2009-2}, potentially allowing for even more complex quantum simulations to be run. 
Our design employs variable RF voltages to temporarily decrease the ion spacing, allowing for an enhanced site-to-site coupling in the ion lattice. For an array of parallel linear traps, such RF shuttling requires only two independent RF voltages, independently of the array size. This is a drastic reduction compared to our previous work with point trap arrays where the number of required RF voltages scales linearly with the array size \cite{Kum2016}. Furthermore, the use of RF voltage tuning allows one to use an ion-electrode separation $d$ that is significantly increased compared to the one required in an array with sufficiently small ion-ion distance and fixed RF voltage. The separation $d$ thereby remains basically constant during RF tuning. This point will be further discussed in section~\ref{sec:concept} and in appendix~\ref{app:advantages_RF_tuning}. Larger separations $d$ can help reduce the motional decoherence rate (ion heating rate), which is a limiting factor in previous two-dimensional ion lattices \cite{Hak2019}.

The article is structured as follows: In section~\ref{sec:concept}, the conceptual design and functionality of the proposed trap array are outlined. In section~\ref{sec:twin-trap}, we consider a minimal instance array consisting of two parallel linear traps with segmented DC electrodes. This minimal instance design possesses the core functionality of the proposed processor, which we show by trap simulations in section~\ref{sec:twin-trap:simulation}. In section~\ref{sec:twin-trap:fab}, we describe the fabrication process and show electrical characteristics of the fabricated trap chips. In section~\ref{sec:twin-trap:measurements}, we demonstrate ion-trapping in multiple trapping sites and characterize a trap chip in terms of electric stray fields and motional heating. Future improvements of the trap design and fabrication, as well as an outlook on future designs are discussed in section~\ref{sec:conclusion}. This discussion is complemented by simulations of a trap array with $10\times10$ trapping sites, shown in appendix~\ref{app:large_array}.

\section{Conceptual design}
\label{sec:concept}

\begin{figure}[b]
	\centering
	\includegraphics[width=0.7\textwidth]{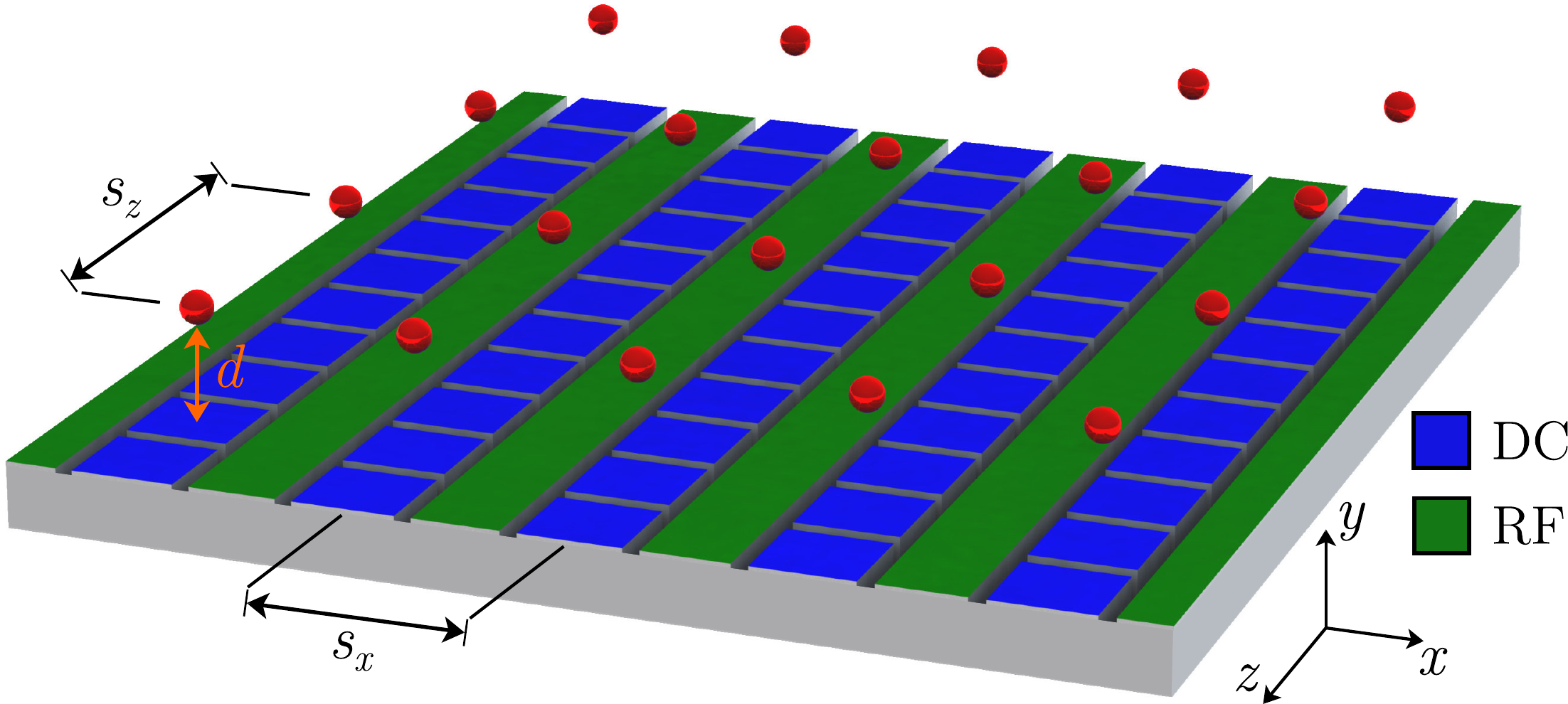}
	\caption{Realization of an ion-lattice quantum processor based on parallel linear traps. Parallel RF rails (green) and segmented DC rails (blue) are used to confine ions (red spheres) in individual trapping sites of a rectangular lattice with trap spacings \trapdistx and \trapdistz. The distance between the ions and the trap surface is $d$.}
	\label{fig:concept_geometry}	
\end{figure}

The electrode geometry of the proposed quantum processor is a linear trap array, as illustrated in figure~\ref{fig:concept_geometry}. 
Colinear rails for radio-frequency (RF) voltage (green) are alternated with segmented rails (blue) that are grounded in the RF domain. This configuration creates parallel linear traps for ions with a spacing \trapdistx at a distance $d$ above the chip surface. Within each linear trap, an additional multiwell potential with periodicity \trapdistz is established along the $z$-direction by applying spatially periodic static voltages (DC) on the blue electrode segments. The combination of RF and DC fields thus creates a two-dimensional lattice of trapping sites with trap spacings \trapdistx and \trapdistz. The quantum states of ions confined within the same trapping site are manipulated using state-of-the-art protocols \cite{Haf2008, Sch2013}. Quantum operations between ions in adjacent trapping sites, such as entangling gates or effective spin-spin interactions, are realized via the coupling of their motional states \cite{Wil2014}. Of central importance for this is the motional coupling strength \cite{Har2011,Wil2014}
\begin{equation}
\label{eq:coupling-rate}
\Wc = \frac{\varsigma_j\,Q^2}{2\pi\epsilon_0 M}\frac{1}{\omega_z\,s_j^3}\,,\text{ with }\varsigma_j=\left\{ \begin{array}{ll}
1 & \textrm{for }j=z\textrm{ (coupling along }z)\\
\frac{1}{2} & \textrm{for }j=x\textrm{ (coupling along }x)\\
\end{array}\right..
\end{equation}
Here $M$ is the mass of the ions in each site, $Q$ their charge, and $\omega_z$ are their (resonant) axial frequencies\footnote{We employ the axial mode for coupling in both directions $x$ and $z$, since it typically has the smallest frequency, $2\wz<\wx,\wy$, yielding the highest coupling rate.}. The motional coupling strength \Wc depends crucially on the trap spacings \trapdistx and \trapdistz, respectively. This has important consequences for the design of the trap array. For the proposed quantum processor, figure~\ref{fig:concept_geometry}, we can envision two distinct design choices: (i) the ion lattice has small trap spacings with sufficient motional coupling \Wc at all times; (ii) the ion lattice has relatively large trap spacings which need to be temporarily decreased to realize an inter-site quantum operation.

Feasible parameters for design choice (i) would be a trap spacing $s_j\approx\SI{40}{\micro\meter}$ with a coupling rate $\Wc\approx2\pi\times\SI{1}{\kilo\hertz}$, as used in \cite{Mie2016}. Furthermore, the coupling strength \Wc between any pair of adjacent trapping sites could be tuned by adjusting the values $s_j$. This can be achieved by controlling the RF and DC voltages in the trap array as explained later on. A downside of design choice (i) is the required small ion-surface separation $d\lesssim s_j$ (see appendix~\ref{app:advantages_RF_tuning} for details). Such a close proximity to the trap surface typically entails a large motional decoherence rate (heating rate) \Gh \cite{Bro2015}. High heating rates $\Gh\gtrsim\Wc$ are a serious impediment for realizing inter-site quantum gates with high fidelity \cite{Wil2014,Hak2019}. While cryogenic trap operation and surface cleaning have been demonstrated to strongly reduce the heating rate, the physical origin of the electric field noise remains unknown and there exists no general procedure that would guarantee a small \Gh \cite{Bro2015}. Even if an ion-surface separation $d\gg s_j$ with a lower \Gh could be realized, for instance by using an electrode geometry different to that in figure~\ref{fig:concept_geometry}, such a trap array would unavoidably have inefficient operating conditions due to the exponential decrease of electrode potentials for distances $d$ much larger than the electrode dimensions \cite{Wes2008}. Indeed, trap arrays typically operate in the regime $d\lesssim s_j$ \cite{Cla2009,Ste2014,Bru2016,Kum2016,Mie2016}.

\begin{figure}[tb]
	\centering
	\includegraphics[width=0.85\textwidth]{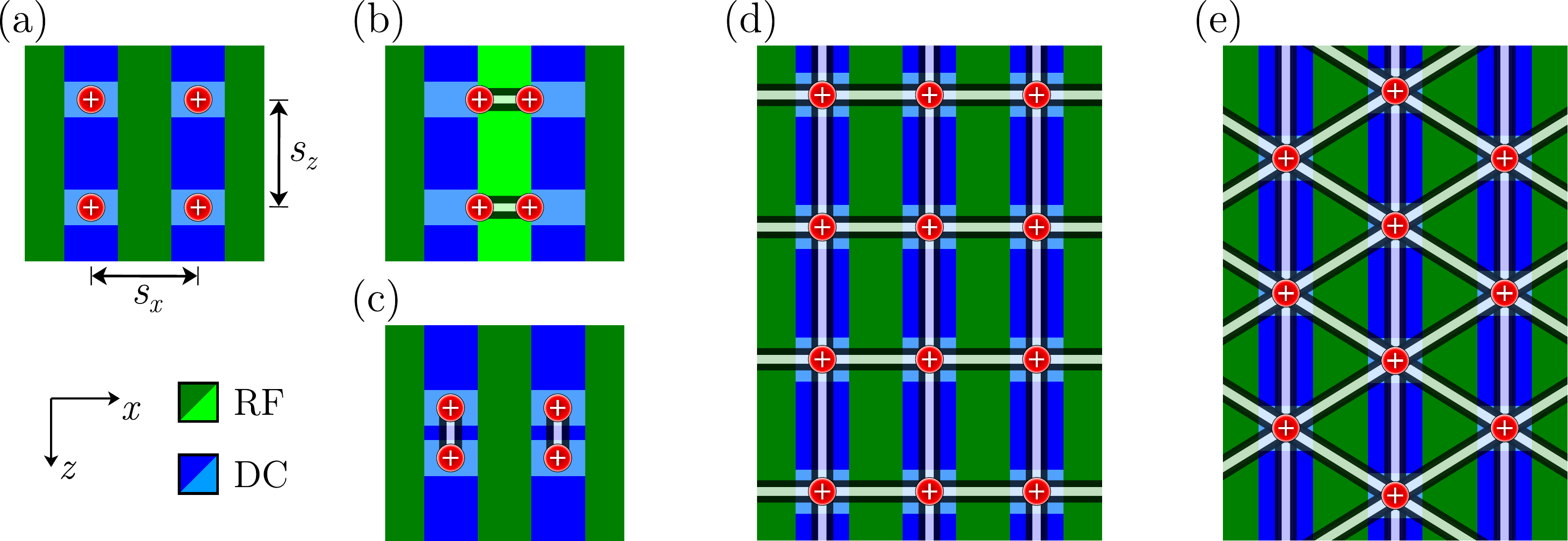}
	\caption{Tunable ion-ion interactions and configurability of the lattice. In the default configuration, ions (red spheres) are stored on a rectangular lattice, (a). The distance \trapdistx between adjacent trapping sites along $x$ is controlled by adjusting the RF voltages, (b). Similarly, the DC voltages are adjusted to control the distance \trapdistz along the $z$-direction, (c). These shuttling operations can be applied to multiple trapping sites simultaneously. A small distance facilitates the creation of entanglement between adjacent ions (white lines). After a sequential application of entangling operations, a nearest neighbor connectivity can be established on the rectangular lattice, (d). Additional shuttling operations of every second column of ions by one lattice period along $z$, with subsequent entangling operations along $x$, allow for the creation of a triangular lattice connectivity, (e). }
	\label{fig:concept_operations}	
\end{figure}

For design choice (ii) one picks modest trap spacings $s_j\sim(\text{100 - 150})\,\si{\micro\meter}$ that allow for a larger electrode-ion separation $d\sim s_j$. The increased distance between ions and trap surface can lead to significantly lower heating rate, since $\Gh\propto d^{-4}$ for many sources of electric field noise \cite{Bro2015}. In the remainder of this article we will therefore focus on design choice (ii). The potential reduction in ion heating rate is traded for having only small motional coupling strengths $\Wc\sim2\pi\times\SI{100}{\hertz}$ between adjacent trapping sites, cf. equation~(\ref{eq:coupling-rate}). Quantum operations between adjacent trapping sites hence need to be realized in a sequential fashion using ion-shuttling operations that temporarily decrease $s_j$. This is illustrated in figure~\ref{fig:concept_operations}. 
Starting from the default configuration, (a), the trap spacing \trapdistx can be strongly reduced by lowering the RF voltage on the RF rail between two adjacent trapping sites (bright green), (b). Similarly, adjacent ions can be brought close along the axial direction $z$ by adjusting the DC voltages on the DC segments, (c). At a reduced distance $s_j=(30\,\textrm{-}\,50)\,\si{\micro\meter}$, coupling strengths $\Wc$ in the \si{\kilo\hertz} range can be achieved, sufficient for coherent operations \cite{Bro2011,Har2011,Wil2014,Hak2019}. The ability to reduce the trap spacing can now be employed to sequentially realize inter-site quantum operations such as entangling gates: Once the desired coupling strength is reached, the secular modes are tuned into resonance and the ions' electronic states are entangled under simultaneous irradiation with laser light, see \eg \cite{Wil2014}. Parallelized entangling operations of pairs of nearest neighbor ions across the array are possible using a global laser field; unwanted coupling between non-nearest neighbors could be drastically reduced by choosing different resonance frequencies for adjacent pairs \cite{Bro2011,Har2011,Wil2014}\footnote{Similarly, tuning a single pair of neighboring ions into resonance effectively suppresses coupling with other ions, and could be used to realize an addressed gate with a global laser field.}, as outlined in appendix~\ref{app:parasitic_coupling}. Subsequently, the lattice spacing is restored to the original value. Entanglement between all nearest neighbors on a rectangular lattice of ions, illustrated in (d), can thus be created using four parallelized shuttling-steps. We emphasize that the ion entanglement, once established, does not depend anymore on the physical arrangement of the ions. This enables the realization of different lattice connectivities using additional ion-shuttling operations along the trap axis $z$. For instance, starting from the connected rectangular lattice, a shift of ions by one lattice spacing along $z$ followed by additional entangling operations along $x$ allows one to establish a triangular lattice connectivity, illustrated in (e). In a similar fashion, $z$-translations of ions can be used to enable entanglement between more distant ions, \eg next-nearest neighbors on the rectangular lattice or even ions at different ends of the lattice. 

The sequential coupling scheme outlined above is useful for various applications. For instance, the motional coupling between ions in adjacent trapping sites could be used for the simulation of spin models in a sequential way (digital quantum simulator) \cite{Lan2011,Wil2014}. Another possibility might be to extend recent studies of entanglement propagation in a linear ion chain \cite{Jur2014} to a two-dimensional ion lattice. Furthermore, one could create cluster states by applying a controlled phase gate (\eg realized by an entangling operation and single qubit rotations) to every pair of neighboring sites in the rectangular ion lattice \cite{Mam2019}. The cluster states could then be used as a resource for a measurement-based quantum processor \cite{Bri2009}. 

In addition to qubit-qubit operations across adjacent trapping sites, the envisioned quantum processor will need single qubit gates, requiring laser-addressing of individual trapping sites. For small and moderate-sized arrays, this could be achieved with a Raman gate \cite{Bal2016}, employing a crossed Raman beam geometry to address a specific ion. Cross talk could be reduced by moving untargeted ions along the trap axes, out of the beam path. For larger arrays, more scalable solutions will eventually be required: integrated optics such as waveguides with Bragg couplers integrated into the trap chip \cite{Meh2016}, or global microwave radiation fields in combination with magnetic field gradients \cite{Pil2014}.

\section{Simplest instance: Linear twin-trap}
\label{sec:twin-trap}

\begin{figure}[htbp]
	\centering
	\includegraphics[width=0.75\textwidth]{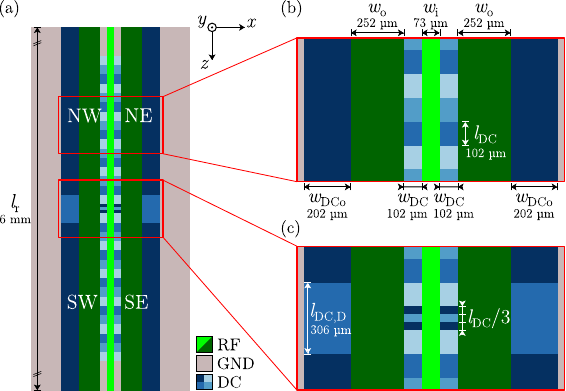}
	\caption{Electrode geometry of the linear twin-trap. (a) Three colinear RF rails (green) with lengths $l_\text{r}=\SI{6}{\milli\meter}$ create radial confinement for two parallel linear traps. The rails have widths of $w_\text{o}=\SI{252}{\micro\meter}$ (outer rails) and $w_\text{i}=\SI{73}{\micro\meter}$ (inner rails). (b) The two DC rails with widths $w_\text{DC}=\SI{102}{\micro\meter}$ are broken up into island-like DC segments of length $l_\text{DC}=\SI{102}{\micro\meter}$. Additional DC rails with width $w_\text{DCo}=\SI{202}{\micro\meter}$ at the edge of the structure are not segmented. The segments in each trap quadrant (NW, NE, SW, SE) are connected periodically to the same voltage supply channel as indicated by the different tones of blue. Electrode voltages in different quadrants can be set independently. (c) At the trap center an axial interaction zone is realized by DC segments with a smaller length of $l_\text{DC}/3$ and an independent segment of the outer DC rails with length $l_\text{DC,D}=\SI{306}{\micro\meter}$.}
	\label{fig:sim_electrode_geometry}	
\end{figure}

The simplest instance of the linear trap array outlined in the previous section is given by two parallel linear traps with segmented DC electrodes, referred to as linear twin-trap in what follows. Figure~\ref{fig:sim_electrode_geometry}\,(a) gives an overview of the electrode geometry. The twin-trap has a central region, surrounded by four identical quadrants (NW, NE, SW, SE). Confinement in the radial ($xy$-) plane is produced by the three RF rails (green), that stretch over the entire length of the trap. In between the RF electrodes there are two segmented DC rails with segment lengths \lDC, (b). The RF and DC rail widths $w_\text{o}, w_\text{i}$ and $w_\text{DC}$ are optimized for maximum trap efficiency , which also leads to close-to-optimal trap depth\footnote{The optimum for the double-well barrier \Ub, defined in figure~\ref{fig:sim_confinement_home}\,(b), coincides with the efficiency optimum; the global trap depth $U_0$ is at about 84\% of its optimal value. See \cite{Hol2020} for more information.}. Within each trap quadrant, the DC segments are periodically connected as indicated by the different tones of blue. This enables the creation of multiwell potentials for axial confinement along the $z$-axis with a lattice spacing $\trapdistz=3\,\lDC$. Additional outer DC electrodes at the edge of the structure (dark blue) are needed to overlap the two DC multiwells with their respective RF null. Within each trap quadrant, the DC segments and outer DC electrodes can be controlled independently.

The periodic assignment of voltages to the DC segments, shown in figure~\ref{fig:sim_electrode_geometry}\,(b), allows one to control the DC multiwells with only 4 DC channels each (3 DC segments and 1 outer DC). Furthermore, the DC multiwells in the left and right linear trap can be independently translated along the $z$-axis, as required for establishing different  lattice connectivities. Ion transport within one linear trap thereby employs the periodicity of the DC segments, similar as in other surface trap designs \cite{Ami2010,Mau2016}. The independent transport in two parallel linear traps relies on the fact that the left multiwell is mainly controlled by DC segments in the left linear trap, while the segments in the right trap have a significantly weaker influence due to the larger spatial separation; and vice versa. This principle should be easily extendable to a larger number of parallel traps.  

The DC islands at the trap center, shown in figure~\ref{fig:sim_electrode_geometry}\,(c), are further split in three segments of length $l_\text{DC}/3$, and also the outer DC rails have an independent segment of length $l_\text{DC,D}$. The finer segmentation in this axial interaction zone allows one to reduce the lattice spacing \trapdistz locally for the central pairs of trapping sites. Alternatively, the three central segments can be treated as one electrode of length \lDC for a seamless transport of the DC multiwells across the central region during axial shuttling operations. 

The twin-trap design, figure~\ref{fig:sim_electrode_geometry}, is similar to the trap used by Tanaka et\,al. \cite{Tan2014}, where parallel ion strings with different RF configurations, leading to different string distances, were demonstrated. We extend that work by adding segmented DC electrodes, which is indispensable for a scalable design and requires multilayer fabrication techniques (described in section~\ref{sec:twin-trap:fab}). The segmentation of the DC rails is also essential to investigate the core functionality of our approach: nearest neighbor interactions within the ion lattice in two spatial dimensions and a configurable lattice connectivity. 
In Ref.~\cite{Tan2014}, different RF amplitude values have been realized using mechanically tuned capacitors. In contrast, the realization of entangling operations in the twin trap design requires to \emph{dynamically} adjust the RF amplitudes. Dynamic control can be achieved using two phase-stabilized RF resonators \cite{Kum2016}; a prototype of such an electrically tunable resonator that can be operated at cryogenic temperatures is described in Ref.~\cite{Hol2020}. We note that the number of required RF resonators does not scale with the size of the trap array (see the section on RF shuttling in appendix~\ref{app:large_array}).

\subsection{Trap simulation}
\label{sec:twin-trap:simulation}

We demonstrate the functionality of the twin-trap design by trap simulations\footnote{For trap simulation, we use the electrode package for Python by R. J{\"o}rdens (\url{https://github.com/nist-ionstorage/electrode}); see also \cite{Sch09,Sch10}.}, considering \Ca ions. First, we analyze the trap confinement in the default configuration of a rectangular lattice with $2\times9$ sites and trap spacings $s_x\approx\SI{105}{\micro\meter}$ and $s_z\approx\SI{306}{\micro\meter}$. Second, we characterize an independent axial translation process where the two adjacent DC multiwells are shifted continuously over one lattice period relative to each other. Such translations are a key requirement for the envisioned configurable lattice connectivity. Third, we simulate tuning of the trap spacings down to values $s_x=\SI{40}{\micro\meter}$ and $s_z=\SI{50}{\micro\meter}$, respectively, giving rise to a motional coupling rate between single ions in adjacent sites of $\Wc\gtrsim2\pi\times\SI{1}{\kilo\hertz}$ in both directions. For all simulated configurations we obtain suitable trapping parameters, \ie secular frequencies on the order of \SI{1}{\mega\hertz}, a trap depth of several tens to hundreds of \si{\milli\electronvolt} and required voltages of $\Urf\approx(100\text{ - }400)\,\si{\volt}$ at RF and a few to a few tens of \si{\volt} DC.

\subsubsection{Default trapping configuration}

\label{sec:twin-trap:simulation:default}
\begin{figure}[htbp]
	\centering
	\includegraphics[width=0.9\textwidth]{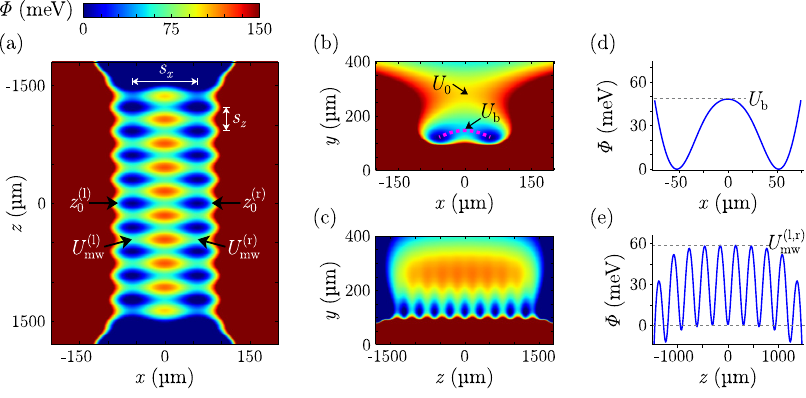}
	\caption{Trap confinement in the default configuration with trap spacings $\trapdistx=\SI{105}{\micro\meter}$ and $\trapdistz=\SI{306}{\micro\meter}$ at an ion-surface separation of $d=\SI{121}{\micro\meter}$. Subplots (a), (b), (c) show cross sections of the total potential $\mathit{\Phi}$ in the $xz$-, $xy$- and $zy$-planes, respectively, crossing the trapping site at $r_0 = (-52.3, 121, 0)\,\si{\micro\meter}$. The color scale is cut off for $\mathit{\Phi}<\SI{0}{\milli\electronvolt}$ and $\mathit{\Phi}>\SI{150}{\milli\electronvolt}$ and we set $\mathit{\Phi}(r_0)=0$. The pink dashed line in (b) illustrates the line of minimal potential between the two central trapping sites at $x_0=\SI{\pm52.3}{\micro\meter}$, $z_0=0$. The potential along this line is shown in (d). (e) Axial multiwell potential at $x_0=\SI{-52.3}{\micro\meter}$.}
	\label{fig:sim_confinement_home}	
\end{figure}

In the default trapping configuration, the inner and outer RF rails are set to the same RF voltage and the DC voltages are applied periodically across all DC segments, with a mirror symmetry between the left and right linear traps. Details on the calculation of DC voltage sets are given in appendix~\ref{app:voltage-sets}. The total confining potential $\mathit{\Phi}$ in this configuration is shown in figure~\ref{fig:sim_confinement_home}.
The potential has 18 individual trapping sites that are arranged in two columns along the two RF nulls, forming a rectangular lattice with trap spacings $\trapdistz=\SI{306}{\micro\meter}$ and $\trapdistx=\SI{105}{\micro\meter}$ (the sites at $|z|\approx\SI{1500}{\micro\meter}$ are not confined). The ion-surface separation is $d\approx\SI{120}{\micro\meter}$. An RF voltage of $\Urf=\SI{142}{\volt}$ at $\Wrf=2\pi\times\SI{23}{\mega\hertz}$ yields a stability factor $q=\wrad \sqrt{8}/\Wrf\approx0.4$, where \wrad is the radial frequency in absence of DC fields (details on the determination of RF parameters are given in appendix~\ref{app:voltage-sets}). The DC voltages for axial confinement are on the order of \SI{1}{\volt} and give rise to an axial frequency $\wz=2\pi\times\SI{1.0}{\mega\hertz}$. The DC confinement leads to a splitting of the radial frequencies, $\wrado, \wradt=2\pi\times(3.1,3.3)\,\si{\mega\hertz}$, and causes a tilt $\thetar=\ang{41.2}$ of the radial modes with respect to the vertical direction $y$. The tilt allows for almost equal laser cooling conditions for both radial modes, assuming laser beam propagation parallel to the trap surface. The axial mode is aligned with the $z$-axis, $\thetaz=0$. The trapping sites are separated from each other by multiwell barriers $\Umwl=\Umwr=\SI{59}{\milli\electronvolt}$ along the $z$-direction and the RF barrier $\Ub=\SI{48}{\milli\electronvolt}$ along the $x$-direction. The barrier $U_0=\SI{102}{\milli\electronvolt}$ defines the global trap depth for ions in radial direction. These trap depths are significantly higher than the depths in other ion-lattice processors \cite{Mie2016}, and allow for an operation of the trap at room temperature. Deviations in the trapping parameters across the lattice due to finite-size effects are relatively small, with variations of the secular frequencies of about \SI{10}{\kilo\hertz} and of the radial mode tilt $\thetar$ by about \ang{5}. The biggest deviations are found at the outermost sites, $z\approx\pm\SI{1230}{\micro\meter}$, where the trap depths $U_0$ and \Umw are reduced by about 25\,\%. The outermost sites are also slightly displaced from the RF null leading to a residual RF electric field $E_\parallel\approx\SI{730}{\volt\per\meter}$ in the laser plane ($xz$). This field causes excess micromotion with a micromotion modulation index $\beta=k z_\text{mm}\approx0.73$ \cite{Ber1998}, where $z_\text{mm}$ is the micromotion amplitude and $k$ is the wavenumber of the $\SI{729}{\nano\meter}$ laser beam driving the $4^{2}\text{S}_{1/2}\leftrightarrow3^{2}\text{D}_{5/2}$ qubit transition in \Ca. For the next inner trapping sites, $z\approx\pm\SI{920}{\micro\meter}$, the shift off the RF null is already notably smaller, with $E_\parallel\approx\SI{270}{\volt\per\meter}$ and $\beta\approx0.27$. These finite size effects result from the finite lengths of the RF rails and the fact that DC fields calculated for the central sites are non-ideal for sites at the trap edges. In future trap designs, such effects could be reduced by increasing the number of independent DC segments and by elongating the RF rails.

\subsubsection{Independent axial translations}

\label{sec:twin-trap:simulation:translations}
\begin{figure}[htbp]
	\centering
	\includegraphics[width=0.9\textwidth]{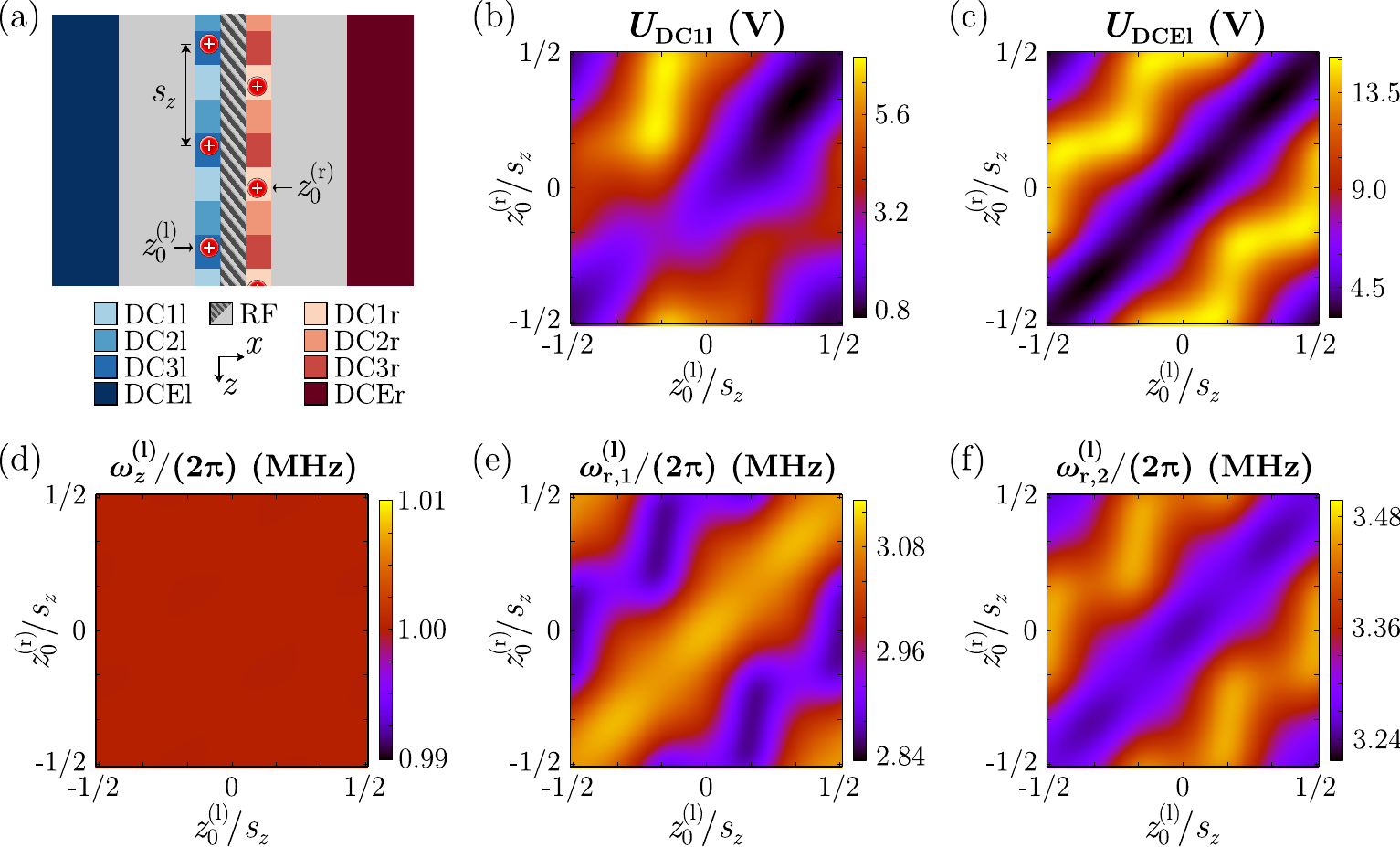}
	\caption{Characterization of independent axial translations. (a) A periodic assignment of DC voltages to the DC segments creates two independent DC multiwells for ions (red spheres), with a lattice spacing of \trapdistz each. Arbitrary positions $z_0^{(l)}, z_0^{(r)}$ of the left and right multiwell can be realized by adjusting the DC voltages. The voltages $U$ required for a nominal axial confinement of $\wz^\text{(l)}= \wz^\text{(r)}=2\pi\times\SI{1}{\mega\hertz}$ are shown in (b) and (c), for segments DC1l and DCEl. The resulting secular frequencies for the trapping site at $z_0^{(l)}$ are shown in (d), (e) and (f).}
	\label{fig:sim_characterization_BBshuttling}	
\end{figure}

One of the goals of the twin trap is to demonstrate a configurable ion lattice connectivity, \eg switch from a rectangular to a triangular lattice. This requires that ions in the left and right linear traps can be moved relative to each other along the trap axis $z$ by at least one lattice period $s_z$. The ions in each linear trap are confined in DC multiwell potentials created by the periodic assignment of voltages to the DC segments. The basic principle of independent axial translations in this setup is illustrated in figure~\ref{fig:sim_characterization_BBshuttling}\,(a).

Grouping the DC electrodes in 8 groups (6 periodically repeating segments and 2 edge electrodes), as indicated by the different segment colors, we calculate voltage sets that simultaneously create axial confinement for two trapping sites at axial positions $z_0^\text{(l)}$ and $z_0^\text{(r)}$ in the left and right RF null, respectively (details in appendix~\ref{app:voltage-sets}). The periodicity of the voltages assigned to the DC segments then creates multiwells with trapping sites at $z_0^\text{(l,r)}\pm m\trapdistz$, where $m=0,1,2,...$ is an integer number. To demonstrate the capability of independent axial translations we analyze the trap parameters obtained for different pairs of multiwell positions across the entire parameter space $(z_0^\text{(l)},z_0^\text{(r)})\in[-\trapdistz/2,+\trapdistz/2]^2$. For each pair of positions, the axial frequencies are set to a constant nominal value $\wz^\text{(l)}= \wz^\text{(r)}=2\pi\times\SI{1}{\mega\hertz}$. Figures~\ref{fig:sim_characterization_BBshuttling}\,(b) and (c) show the DC voltages required to realize the multiwells at positions $(z_0^\text{(l)},z_0^\text{(r)})\in[-\trapdistz/2,+\trapdistz/2]^2$. These voltages, displayed for electrodes DC1l and DCEl, are on the order of (1 - 10)\,\si{\volt}\footnote{The voltages for segments DC2l, DC3l can be obtained by shifting $z_0^\text{(l,r)}$ in (b) by $\lDC=\trapdistz/3$ (translational symmetry of the trap). The voltages for the segments on the right side, DC1r to DCEr, are given by swapping the axes $z_0^\text{(l)}$ and $z_0^\text{(r)}$ (mirror symmetry).}. The axial frequency, shown in (d), maintains the nominal value $\wz=2\pi\times\SI{1}{\mega\hertz}$ with high accuracy for all pairs of positions. The radial modes, (e) and (f), show a variation of $\sim 10\,\%$ across the full parameter space\footnote{The data shown in figure~\ref{fig:sim_characterization_BBshuttling}\,(d)-(f) are for the site at $z_0^{(l)}$. The frequencies for the site at $z_0^{(r)}$ are given by swapping the axes $z_0^\text{(l)}$ and $z_0^\text{(r)}$.}. 
Other trap parameters (not shown) show slight variations as well. For instance, the radial mode tilt relative to the $y$-axis varies between $\thetar\approx\ang{30}\textrm{ - }\ang{40}$. For the trap depths, values $\Ub,\Umw>\SI{48}{\milli\electronvolt}$ and $U_0>\SI{98}{\milli\electronvolt}$ are maintained, similar to the default configuration. More information is given in \cite{Hol2020}. We note that one can choose the axial frequencies $\wz^\text{(l)}, \wz^\text{(r)}$ independently, even to the point that one multiwell is switched off. However, trap depths are maximized when both multiwells are operated with similar \wz.

Any trajectory through the simulated parameter space $(z_0^\text{(l)},z_0^\text{(r)})\in [-\trapdistz/2,+\trapdistz/2]^2$ corresponds to a specific axial translation process. The ability to maintain the multiwell confinement for the entire parameter space demonstrates that translation processes with arbitrary multiwell positions are possible. Furthermore, the simulation of such a wide range of control parameters has the advantage that promising parameter space trajectories, for instance those with a minimal variation in secular frequency or mode tilt, can be quickly identified. However, the approach does not deliver a time-dependent voltage sequence that implements a specific temporal dependence $z_0^\text{(l)}(t), z_0^\text{(r)}(t)$ of the well positions. Such voltage sequences can be engineered in various ways. Typically, the aim is to maintain low motional excitation during the shuttling (adiabatic transport) \cite{Row2002,Bla2011} or to cancel excitations at the end of the sequence (diabatic transport) \cite{Bow2012,Wal2012}. The full parameter scan presented here may serve as a starting point for the calculation of such sequences.

We emphasize that the grouping of DC segments significantly reduces the required number of DC control voltages for axial translation processes. In the present design, only 8 control voltages are needed to independently move the two multiwells over arbitrary distances. Other adjustments of the trapping potential can be realized using additional groups of segments, foremost the independent segments in the axial interaction zone that allow one to reduce the trap spacing $s_z$ (see next section). In future designs one could add even more DC segments to improve on the control of the trapping potential at individual sites, \eg for micromotion compensation and secular frequency adjustments.

\subsubsection{Adjustment of trap spacings}

\label{sec:twin-trap:simulation:lattice-constants}
\begin{figure}[htbp]
	\centering
	\includegraphics[width=0.9\textwidth]{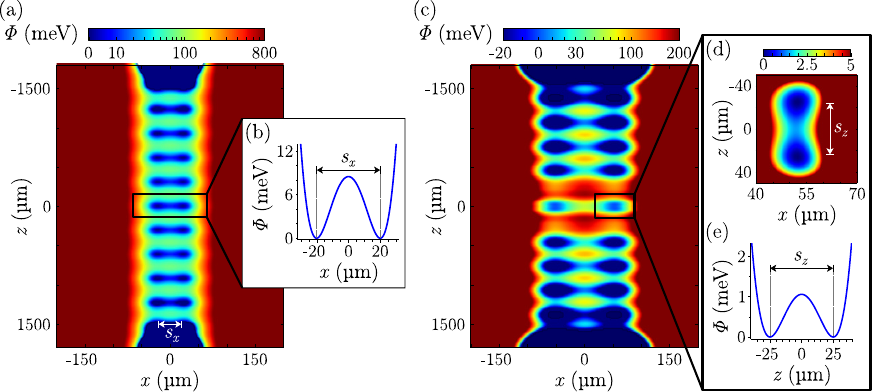}
	\caption{Confinement for enhanced ion-ion interaction strengths along the two lattice directions. 
		(a) Reduced RF configuration with $\trapdistx\approx\SI{40}{\micro\meter}$. The cross section of the total potential $\mathit{\Phi}$ in the $xz$-plane crosses the trapping site at $r_0 = (-20.1, 121, 0)\,\si{\micro\meter}$. The color scale is non-linear for better visibility of the minima and we set $\mathit{\Phi}(r_0)=0$. The inset, (b), shows the double-well potential across the RF barrier between the two central trapping sites at $x_0=\SI{\pm20.1}{\micro\meter}$, $z_0=0$. (c) Configuration with reduced axial distance $\trapdistz\approx\SI{50}{\micro\meter}$ in the axial interaction zone at the trap center. The cross section of the $xz$-plane crosses the trapping site at $r_0^\text{(c)} = (-52, 121, 25)\,\si{\micro\meter}$. The insets, (d) and (e), show a zoom-in on the central double-well potential that is formed along the trap axis $z$.}
	\label{fig:sim_confinement_interaction}	
\end{figure}

The creation of entanglement between ions in adjacent lattice sites requires a reduction of the trap spacings to enhance the coupling rate \Wc. Along the $x$-direction, the trap spacing $s_x$ is reduced by attenuating the RF voltage $\Urf^\textrm{(i)}$ on the inner RF rail relative to the voltage $\Urf^\textrm{(o)}$ on the outer RF rails. Figure~\ref{fig:sim_confinement_interaction} shows the trapping potential in such an ``attenuated RF'' configuration for a reduced trap spacing $s_x=\SI{40}{\micro\meter}$. The axial multiwell confinement is preserved for all 18 trapping sites with an axial frequency $\wz=2\pi\times\SI{1.0}{\mega\hertz}$. The corresponding motional coupling rate for two \Ca ions in adjacent trapping sites across the RF barrier is $\Wc=2\pi\times\SI{1.4}{\kilo\hertz}$, cf.~equation~(\ref{eq:coupling-rate}). The RF double well potential, shown in the inset (b), is well defined with a radial barrier of $\Ub=\SI{8.5}{\milli\electronvolt}$\footnote{We note that the double-well potentials in figures~\ref{fig:sim_confinement_interaction}\,(b) and (e) are well described by a model potential of the form $\mathit{\Phi}_\text{dw}(\zeta)=a \zeta^4 - b \zeta^2$. Fixing the well distance $s_\zeta=\sqrt{2b/a}$ and the secular frequency $\omega_\zeta=\sqrt{4b/M}$, the expected barrier is $U_\text{b,dw}=M \omega_\zeta^2 s_\zeta^2/32$. For the potentials in figures~\ref{fig:sim_confinement_interaction}\,(b) and (e) one finds $U_\text{b,dw}=\SI{8.9}{\milli\electronvolt}$ and $U_\text{b,dw}=\SI{1.1}{\milli\electronvolt}$, respectively, in good agreement with the simulated barriers.}. The required RF voltages in this configuration are $\Urf^\textrm{(i)}=\SI{296}{\volt}$ and $\Urf^\textrm{(o)}=\SI{372}{\volt}$ with a stability factor $q=0.4$, identical to the default configuration. The increase of the voltage $\Urf^\textrm{(o)}$ on the outer rails, required by the decreased efficiency of the trap, significantly improves the trap depth to $U_0=\SI{702}{\milli\electronvolt}$. Other trap parameters are similar to the default configuration. The radial frequencies are $\wrado, \wradt=2\pi\times(3.1,3.3)\,\si{\mega\hertz}$, the multiwell barrier is $\Umw\approx\SI{60}{\milli\electronvolt}$. The axial mode remains aligned with the $z$-axis, $\thetaz=0$, and the radial mode tilt is $\thetar=\ang{10.2}$. The DC voltages required to sustain the axial multiwell potential remain on the order of \SI{1}{\volt}. We note that the trap spacing $s_x$ slightly differs along the trap axis, with values $s_x=\SI{40}{\micro\meter}$ at the trap center, $z=0$, and $s_x\approx\SI{43}{\micro\meter}$ at the outermost sites, $z\approx\pm\SI{1200}{\micro\meter}$. The difference in trap spacing is caused by finite size effects in the trap and leads to a variation in coupling strength of about $\Delta\Wc\approx2\pi\times\SI{0.3}{\kilo\hertz}$. The finite size effects could be decreased in future designs (see section~\ref{sec:twin-trap:simulation:default}).

Along the axial direction, the trap spacing $s_z$ can be reduced in the axial interaction zone at the trap center where the DC island electrodes have a finer segmentation. Ions outside the interaction zone remain in a periodic DC multiwell potential as in the default configuration. Fig.\,\ref{fig:sim_confinement_interaction}\,(c) shows the confining potential for a configuration where the axial distance in the interaction zone is reduced to $s_z=\SI{50}{\micro\meter}$; (d) shows a magnified view of the 2 central sites forming a double well. These sites have radial frequencies $\wrado, \wradt=2\pi\times(3.1,3.3)\,\si{\mega\hertz}$ identical to the default configuration. The axial mode has a frequency $\wz=2\pi\times\SI{0.91}{\mega\hertz}$ and is tilted by $\thetaz=\ang{8.0}$ relative to the $z$-axis (currently, \thetaz is an unconstrained parameter, which could be improved in future designs by adding additional DC electrodes). The central sites are separated from each other by an axial double well barrier $\Ub^\text{(ax)}=\SI{1.1}{\milli\electronvolt}$, shown in (e). The expected motional coupling between single \Ca ions in these sites is $\Wc=2\pi\times\SI{1.5}{\kilo\hertz}$, cf.~equation~(\ref{eq:coupling-rate}). The axial frequencies in the two central sites can be tuned independently. Micromotion compensation, however, is limited to shifting both sites simultaneously due to the small axial separation $s_z=\SI{50}{\micro\meter}$. On the other hand, given that $s_z$ is substantially smaller than the ion-surface distance $d=\SI{120}{\micro\meter}$, stray fields should be relatively homogeneous across the two sites. Due to the condition $s_z<d$, the double-well potential in the axial interaction zone is not created efficiently, and up to \SI{34}{\volt} must be applied to the central DC segments. The outer 16 trapping sites, $|z_0^\text{(o)}|\gtrsim\pm\SI{459}{\micro\meter}$, are maintained by the periodically connected DC segments with trapping parameters similar to the default configuration. We note that the configuration with reduced axial distance, figure~\ref{fig:sim_confinement_interaction}\,(c), can be seamlessly transformed to the default configuration in figure~\ref{fig:sim_confinement_home} using a two-stage shuttling process. In the first step, the initial separation $s_z=\SI{50}{\micro\meter}$ between the innermost sites is increased to $\SI{306}{\micro\meter}=3\lDC$, realizing a multiwell configuration with constant lattice spacing across the entire length of the chip. The second step then uses an axial translation of the ion lattice to shift the central multiwell site into the origin at $z=0$.

\subsection{Trap fabrication}
\label{sec:twin-trap:fab}

The linear twin-trap design requires multiple metal layers and vertical interconnect access (via) due to the presence of island-like electrodes. The fabrication is carried out at the industrial facilities of Infineon Technologies in Villach, Austria. In general, our fabrication is similar to the CMOS foundry processes recently used for ion traps \cite{Meh2014}. However, while typical CMOS processes are set up for low-voltage logic applications, our processes are optimized for high power and high current applications more suited for ions traps. We also employ a dedicated workstream for the trap fabrication and are therefore not affected by the requirements of other technologies on the same wafer. Established design rules, continuous process monitoring, inline testing and analysis capabilities provide high precision and reproducibility of the devices. For the fabrication of a prototype version of the linear twin-trap, 90 process steps were applied on top of a \SI{725}{\micro\meter} thick silicon substrate\footnote{Boron-doped, room-temperature resistivity $\rho=\SI{3}{\ohm\centi\meter}$} to produce six main functional layers as sketched in figure~\ref{fig:fab_process}:
\begin{figure}[htbp]
	\centering
	\includegraphics[width=0.8\textwidth]{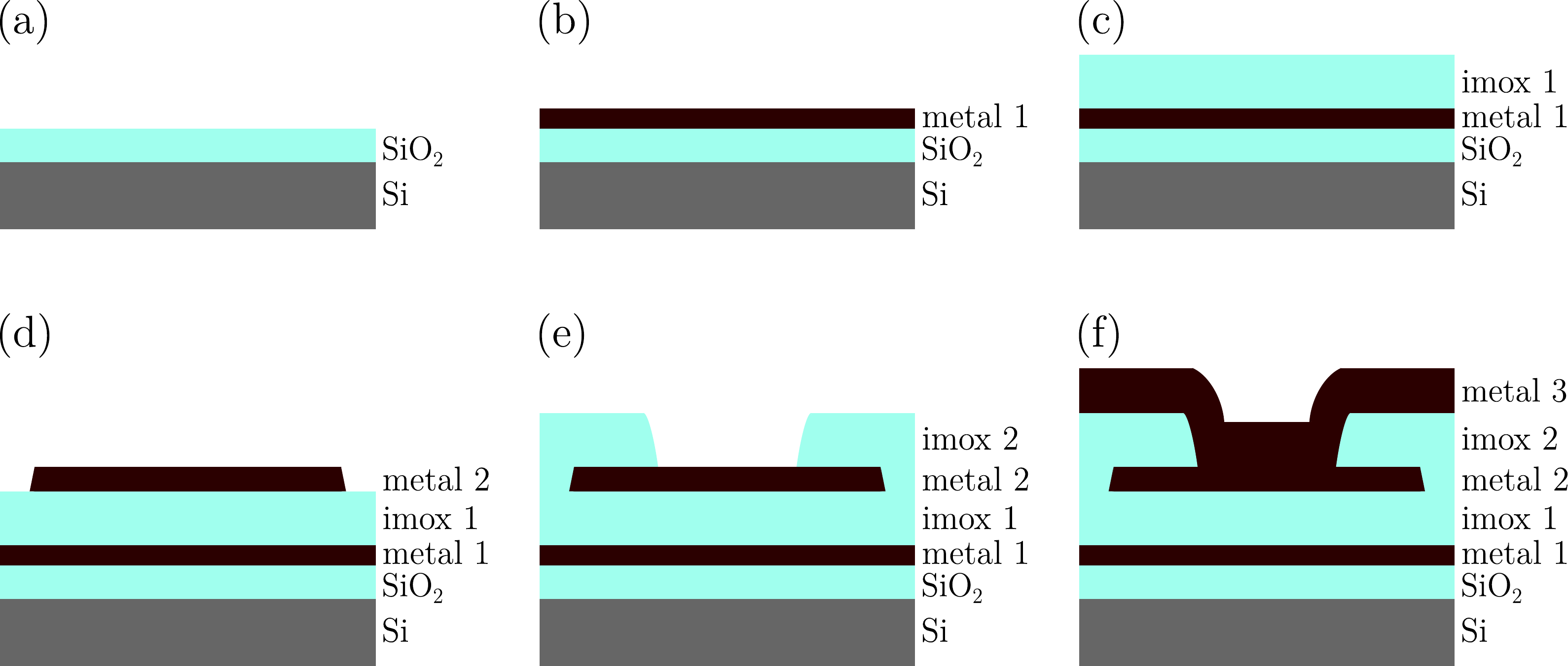}
	\caption{Main fabrication steps, shown for a vertical interconnect access (via) between metal layers 2 and 3. (a) Thermal oxidation of the silicon substrate. Deposition of (b) metal 1, (c) inter-metal-oxide (imox) 1, (d) metal 2, (e) imox 2, and (f) metal 3. All layers are structured by optical lithography and subsequent etching.}
	\label{fig:fab_process}	
\end{figure}
First, a \SI{1300}{\nano\meter} thick SiO$_{2}$ layer is created by thermal oxidation of the Si substrate. This bottom oxide has low defect density and low interface roughness and serves as electrical insulation between substrate and the metal 1 layer. Subsequently, three metal layers are deposited, separated by two \SI{2200}{\nano\meter} thick inter-metal oxide layers (imox). The \SI{750}{\nano\meter} thick metal 1 layer provides (i) shielding of the substrate from RF fields and lasers \cite{Meh2014}, and (ii) shielding of the ion from charge fluctuations in the substrate. The \SI{1000}{\nano\meter} thick metal 2 layer is mainly used for routing of the island-like electrodes to the bonding pads. Metal 3 has a thickness of \SI{2000}{\nano\meter} and defines the trap electrodes. 

All metal layers are made from AlSiCu, an alloy consisting mainly of aluminium. \SI{1}{\percent} silicon and \SI{0.5}{\percent} copper are included to suppress eutectic mixing with the silicon substrate and to increase the resilience to high currents, respectively. The metalization for electrodes and routing has to be low-Ohmic in order to minimize RF pickup voltages on the DC electrodes, to minimize Johnson noise and to minimize heating of the RF rails by capacitive loading currents during trap operation. The imox layers consist of SiO$_{x}$, $x\approx2$, created by low-temperature plasma deposition since the thermal budget of AlSiCu is limited to a maximum temperature  $T_{\mathrm{max}}\approx\SI{400}{\celsius}$. 

Standard optical lithography followed by etching is performed to define the structures within each layer. Vias between the metal layers are defined by etching a funnel-shaped aperture into the separating imox layer which guarantees reliable coverage of the vias' sidewalls by the upper metal. The structuring of the imox layers is optimized using a focus exposure matrix. In order to guarantee process stability, in-line data of layer thicknesses, critical dimensions, reflectivities and overlay precision are measured and recorded automatically. 

\begin{figure}[b]
	\centering
	\includegraphics[width=0.5\textwidth]{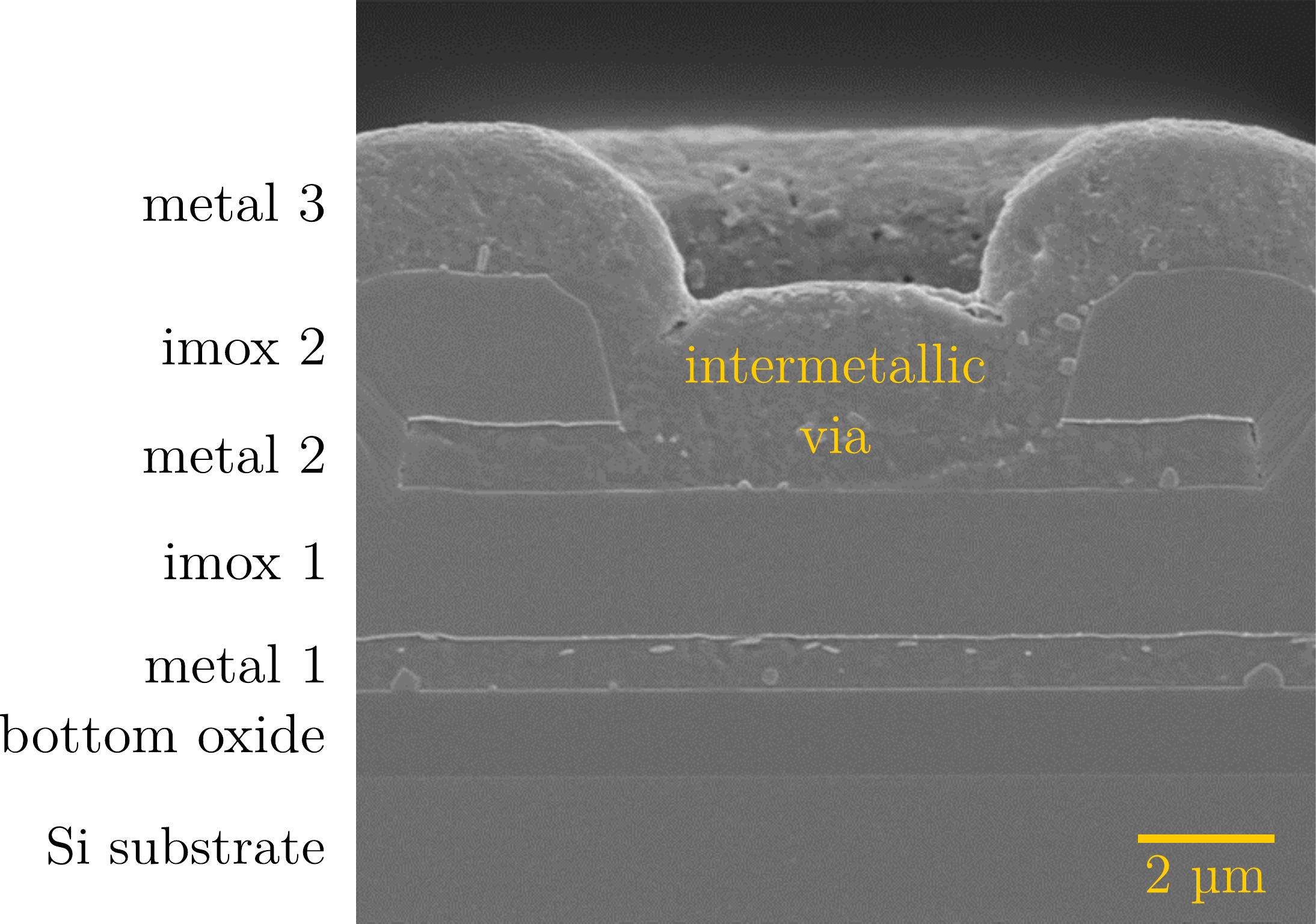}
	\caption{Scanning electron micrograph of a trap cross section showing the metallic and dielectric layers with a via between metal 2 and 3. Material interfaces have been highlighted by HF decoration etching. The aperture in imox 2 is funnel-shaped to improve the sidewall coverage by metal 3.}
	\label{fig:fab_layers}	
\end{figure}

After mechanical dicing into individual chips, electrical analysis (resistance and DC breakdown measurements at room temperature and $T \approx \SI{20}{\kelvin}$) as well as physical analysis (inspection of cross sections) are performed for quality control. The cross section of a via between metal 2 and 3 is shown in the scanning electron microscope (SEM) image in figure~\ref{fig:fab_layers}.
In order to provide high material contrast, the sample has been cut and polished followed by a $\SI{10}{\second}$ exposure to hydrofluoric acid which etches a few nanometers of SiO$_{x}$ and emphasizes the material boundary of SiO$_{x}$. Finally, the sample is sputter-coated with about $\SI{2}{\nano\meter}$ of palladium to maximize the total contrast in the SEM image. The cross section confirms the reliable via connection between metal 2 and 3.

\begin{figure}[b]
	\centering
	\includegraphics[width=0.85\textwidth]{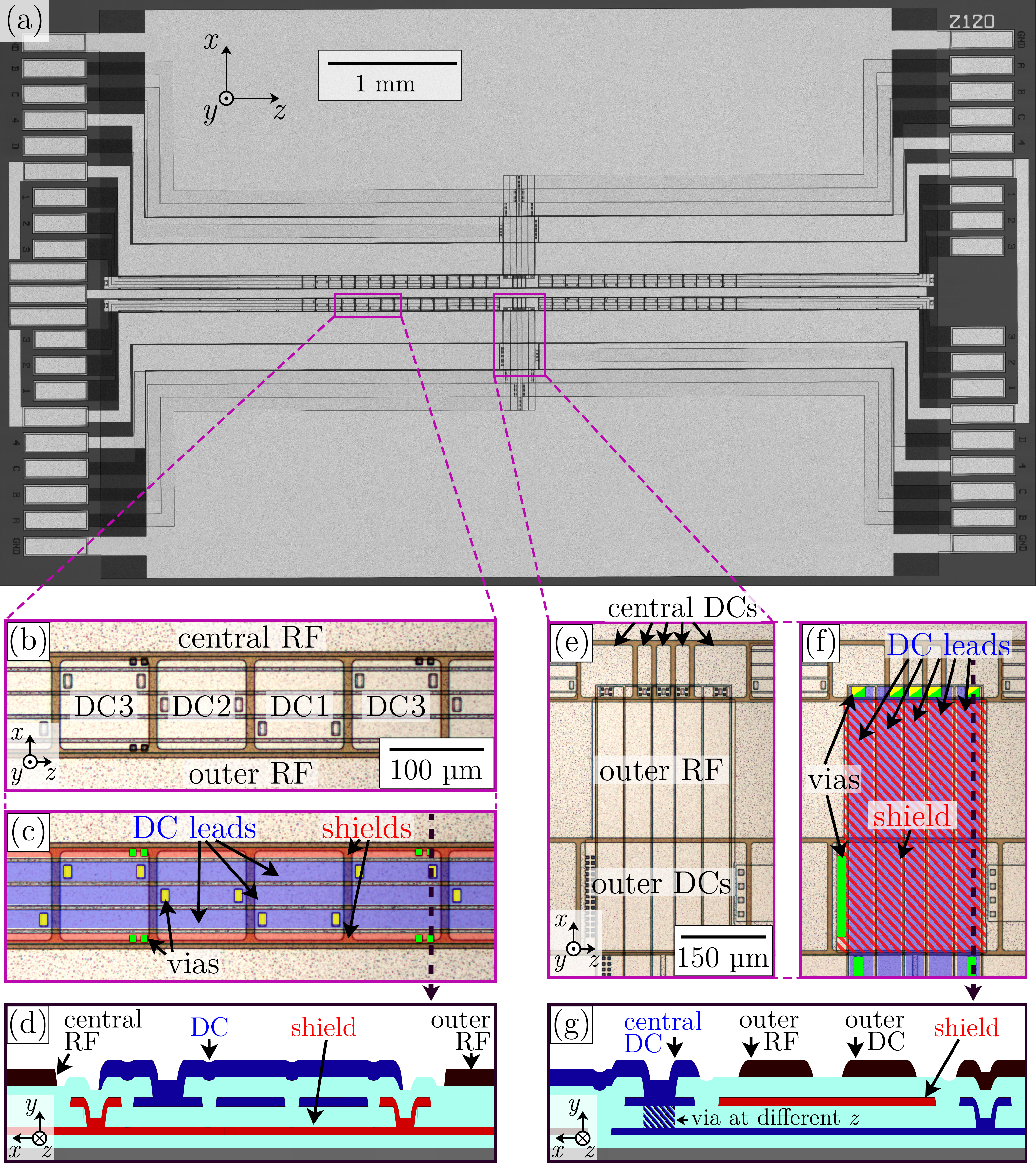}
	\caption{(a) Microscope image of the prototype device. (b) Magnified view of the island-like electrode segments. (c) The same view with a color overlay. Leads on metal 2 (blue) connect to the DC segments with two vias (yellow) per segment. Additional shield leads (red) on metal 2 reduce the parasitic capacitance between the segment leads and the adjacent RF electrodes. The shields are connected to the metal 1 ground with vias (green). The dashed arrow marks the $z$-position of the schematic cross section shown in (d). (e) Magnified view of the trap center with the axial interaction zone. (f) The same view with a color overlay. The central DC electrodes are routed underneath the RF rail through metal 1 leads (blue). A grounded shield (hatched red) on metal 2 reduces the parasitic capacitance between the DC leads and the RF rail. The dashed arrow marks the $z$-position of the schematic cross section shown in (g).}
	\label{fig:fab_traps}	
\end{figure}

A microscope image of the full prototype device is shown in figure~\ref{fig:fab_traps}\,(a). 
The 80 trap electrodes (bright) in the metal 3 layer are separated by \SI{9}{\micro\meter}-wide gaps (dark) and are connected to the bonding pads on the left and right sides of the chip. Vias and traces in lower layers are visible due to the surface topology of the chip. Figure~\ref{fig:fab_traps}\,(b) shows a magnified view of the island-like DC segments DC1, DC2, DC3 in one of the trap quadrants. Every third segment is connected to the same lead on metal 2, as required for the creation of DC multiwell potentials and axial translations. The color code in (c) and in the cross section (d) illustrates the routing to the DC segments on metal 2. RF-pickup on the segments is minimized by two measures: First, vias at both ends of each segment reduce the lead resistance $R_\text{lead}$ since the metal 2 and metal 3 layers are routed in parallel. Within the segmented rail region, the calculated reduction of the lead resistance is about 27\%. Second, additional shield electrodes connected to GND reduce the parasitic coupling capacitance $C_\textrm{p}$ between the DC segments and the adjacent RF electrodes. We quantify the shielding with finite element simulations of a trap cross section: the presence of the metal 1 GND layer reduces $C_\textrm{p}$ by about 92\%; the grounded shields on metal 2 lead to an additional reduction of about 19\%. To further minimize the coupling capacitance, crossing of DC leads below the RF rails is avoided whenever possible. Figure~\ref{fig:fab_traps}\,(e) shows a magnified view of the axial interaction zone at the trap center. The routing to the individually connected central DC segments has to cross the RF rails, as shown in (f) and (g). Therefore, the routing is moved to the metal 1 layer to make room for a shield electrode on metal 2. This minimizes RF-pickup on the lines while maintaining the screening of the Si substrate from laser light. From finite element simulations, we estimate a parasitic coupling capacitance to the RF rails of $C_\textrm{p}\lesssim\SI{0.01}{\pico\farad}$ for any DC electrode. 

The trap chips are produced on wafers with a diameter of \SI{200}{\milli\meter} (8''), holding more than 700 chips. Multiple trap geometries are fabricated simultaneously. In addition to the design with ion-surface separation $d=\SI{120}{\micro\meter}$ described in this article, a slightly adapted geometry with $d=\SI{80}{\micro\meter}$ is on the wafer. Additionally, both electrode geometries are realized in two versions. In one version metal 1 is unstructured, apart for the routing to the central DC segments. The continuous metal 1 layer ensures shielding of the substrate from laser light and reduces the penetration of RF fields into the substrate \cite{Meh2014}. In a second version, about 79\% of the metal 1 layer is removed below the RF electrodes. This trades substrate shielding for a lower capacitance of the RF lines ($\approx\SI{11}{\pico\farad}$ instead of $\approx\SI{29}{\pico\farad}$, estimated from a parallel plate capacitor model), allowing for a larger voltage gain of a step-up resonator \cite{Bra2014} provided the substrate has negligible RF loss \cite{Nie2014}\footnote{We note that in the absence of a shield layer, light-induced charge carrier generation can severely increase the RF losses in the substrate. We have observed such an effect for traps from a different fabrication run, where a high-resistivity Si substrate ($\rho>\SI{10}{\kilo\ohm\centi\meter}$) was used. In addition, light-induced charge carriers can also lead to micromotion that cannot be compensated \cite{Lak2019}.}. 

The wafer layout also contains structures dedicated to the electrical testing of the resistivity of the metal layers and of the resistances of the intermetallic vias. These quantities are used to estimate the amount of RF pickup and Johnson noise on the trap electrodes. The layer resistivities and via resistances are determined in a 4-wire measurement at $T \approx \SI{20}{\kelvin}$\footnote{Layer resistivities are calculated from a resistance measurement of a \SI{65.5}{\milli\meter} long meander trace with \SI{14}{\micro\meter} width, realized on each of the metal layers. Via resistances are calculated from the resistance of 30 identical vias connected in series.}. The results are listed in tables~\ref{tab:tests:resistivity} and \ref{tab:tests:vias}. 
\begin{table}[htbp]
	\parbox{.49\linewidth}
	{
		\centering
		\caption{Resistivity $\rho$ of the three metal layers at $T \approx \SI{20}{\kelvin}$.}
		\label{tab:tests:resistivity}	
		\begin{threeparttable}
			\begin{tabular}{ccc}
				\toprule
				layer & & $\rho$ (\si{\ohm\meter}) \\
				\midrule\midrule
				metal 3 & & \num{2.41\pm0.03e-9}\\
				\midrule
				metal 2 & & \num{2.58\pm0.02e-9} \\
				\midrule
				metal 1 & & \num{2.54\pm0.01e-9} \\
				\bottomrule
			\end{tabular}
		\end{threeparttable}
	}
	\hfill
	\parbox{.45\linewidth}{
		\centering
		\caption{Intermetallic via resistances $R_\text{via}$ at $T \approx \SI{20}{\kelvin}$.}
		\label{tab:tests:vias}	
		\begin{threeparttable}
			\begin{tabular}{ccc}
				\toprule
				via type & & $R_\text{via}$ (\si{\milli\ohm})\\
				\midrule\midrule
				metal 3 to metal 2 & & \num{2.70\pm0.02}\\
				\midrule
				metal 2 to metal 1 & & \num{6.52\pm0.03}\\
				\midrule
				metal 3 to metal 1 & & \num{4.94\pm0.04}\\
				\bottomrule
			\end{tabular}
		\end{threeparttable}
	}
\end{table}
We find reproducible values of the AlSiCu bulk resistivity of $\rho\approx\text{(2.4 - 2.6)}\times\SI{e-9}{\ohm\meter}$ which is comparable to other low-resistivity alloys of aluminium at $T = \SI{20}{\kelvin}$ \cite{Cla1970}. The via resistance depends on the aspect ratio of the metallized via sidewalls and the distance between the connected metal layers. Vias connecting metal 1 to metal 3 are realized with one metal 2 to metal 3 via and two metal 1 to metal 2 vias in parallel to reduce the resistance. All via resistances are on the order of a few \si{\milli\ohm}, demonstrating a good electrical connection across metal layers\footnote{At room temperature, the layer resistivities and the via resistances are both about 10 times larger than at $T = \SI{20}{\kelvin}$. The resistivity measured at room temperature agrees with the literature value for the resistivity of aluminium $\rho_\text{Al}=\SI{2.7e-8}{\ohm\meter}$ \cite{Lide1994}.}.

From the measured resistances we estimate the amount of Johnson noise on the trap electrodes and the corresponding heating rate for a trapped ion (details in appendix~\ref{app:field-noise}). The dominant contribution to the Johnson noise seen by an ion comes from the metal 2 leads for the periodically
connected DC segments. These leads have a resistance of $R_\mathrm{lead} \approx\SI{0.46}{\Omega}$ at $T \approx \SI{20}{\kelvin}$, the via resistances can be neglected. The axial ion heating rate caused by Johnson noise across $R_\mathrm{lead}$ is $\Gh^\textrm{(JN)}\approx \SI{0.015}{phonons\per\second}$, calculated for a \Ca ion with an axial frequency of $\wz=2\pi\times\SI{1}{\mega\hertz}$ (in radial direction the heating rate is on the same order of magnitude). Such a low heating rate is negligible for all practical purposes. 

The RF pickup voltage $U_\textrm{p}$ on the DC electrodes is estimated from an electrical model, considering the electrodes' grounding in the RF domain (details in appendix~\ref{app:field-noise}). Large pickup voltages can induce significant RF electric fields at the ion position which in turn result in excess micromotion that cannot be compensated \cite{Ber1998}. We estimate a very small amount of RF pickup $|\epsilon_\textrm{p}|=|U_\textrm{p}/\Urf|\approx\num{1e-6}$, where \Urf is the applied RF voltage at a frequency $\Wrf=2\pi\times\SI{25}{\mega\hertz}$. Excess micromotion from the corresponding RF electric fields should therefore be negligible.

The maximum required RF voltages on the trap are about $\Urf=\SI{400}{\volt}$, needed in the configuration with reduced trap spacing \trapdistx between the two linear traps (cf. section~\ref{sec:twin-trap:simulation:lattice-constants}). For a reliable trap operation, the dielectric imox layers need to withstand such RF voltages without electrical breakdown. We measure the DC dielectric breakdown voltage between metal layers 2 and 3 directly on the $d=\SI{120}{\micro\meter}$ prototype chips at room temperature and in vacuum. From a set of 10 devices, we observe dielectric breakdown voltages of $\SI{800}{\volt}<V_{\mathrm{BD}}<\SI{1000}{\volt}$. This is in reasonable agreement with the typical dielectric strength \SI{5.6}{\mega\volt\per\centi\meter} of sputter-deposited SiO$_2$ \cite{Bar2009}, given the $\approx \SI{2}{\micro\meter}$ thickness of the imox layers. Furthermore, the measured $V_{\mathrm{BD}}$ is well above the required voltage of \SI{400}{\volt} assuming similar dielectric breakdown mechanisms for DC and RF.

\subsection{Trap characterization}
\label{sec:twin-trap:measurements}

We have performed tests of the fabricated linear-twin traps by means of ion measurements with \Ca ions. The tests include trapping and axial translations of multiple ions, as well as a characterization of stray electric fields and heating rates. The experiments are performed in a closed-cycle cryostat with a base temperature of $T\approx\SI{10}{\kelvin}$ \cite{Nie2015}, while the ion trap is at an operation temperature of $T\approx\SI{50}{\kelvin}$. The elevated temperature of the trap is due to RF absorption in the Si substrate at the location of the RF rails' bonding pads, where there is no grounded shield layer on metal 1. This heating effect could be significantly reduced in future designs by extending the metal 1 shield layer to the bonding pads, thereby inhibiting the RF field penetration into the substrate while adding only slightly to the trap capacitance. \Ca ions are produced from a neutral atom flux by a two-step photoionization process using overlapped laser beams at \SI{422}{\nano\meter} and \SI{379}{\nano\meter} wavelength. In order to cool the ions into the motional ground state we use Doppler and resolved sideband cooling techniques \cite{Lei03}. The $4^{2}\text{S}_{1/2}\leftrightarrow4^{2}\text{P}_{1/2}$ dipole transition at \SI{397}{\nano\meter} is used for Doppler cooling and detection. The $4^{2}\text{S}_{1/2}\leftrightarrow3^{2}\text{D}_{5/2}$ quadrupole transition at \SI{729}{\nano\meter} is used for resolved-sideband operations and spectroscopy. Additional lasers at \SI{866}{\nano\meter} and \SI{854}{\nano\meter} are employed to repump population from the D states back to the P levels. The \SI{397}{\nano\meter}, \SI{866}{\nano\meter} and \SI{854}{\nano\meter} beams are shaped by a set of cylindrical lenses to obtain highly-elliptical beams with a beam waist $w_0 \approx \SI{900}{\micro\meter}$ in the horizontal plane and (20 - 30)\,\si{\micro\meter} in the vertical direction. These elliptical beams are used to cool and image ions in multiple lattice sites simultaneously, as well as during ion shuttling. All other laser beams are circular and address a single trapping site at a time. For trap operation, we apply an RF amplitude $\Urf\approx\SI{180}{\volt}$ at \SI{25}{\mega\hertz} to all three RF rails, resulting in radial frequencies $\wrad\approx2\pi\times\text{(2 - 3)}\,\si{\mega\hertz}$. Axial multiwell confinement with $\wz\approx2\pi\times\SI{1}{\mega\hertz}$ is achieved by applying DC voltages on the order of \SI{1}{\volt}, using the segment connectivity shown in figure~\ref{fig:sim_characterization_BBshuttling}\,(a): The periodic assignment of voltages to the DC segments is extended across the entire length of the trap chip to allow for seamless axial translations of ions in the left and right linear trap. However, due to a short in one of the cables of the cryostat, electrodes DC2l and DC2r had to be connected to the same supply line. Thus, the freedom of moving the two chains independently was limited in the experiments.

\begin{figure}[htbp]
	\centering
	\includegraphics[width=0.9\textwidth]{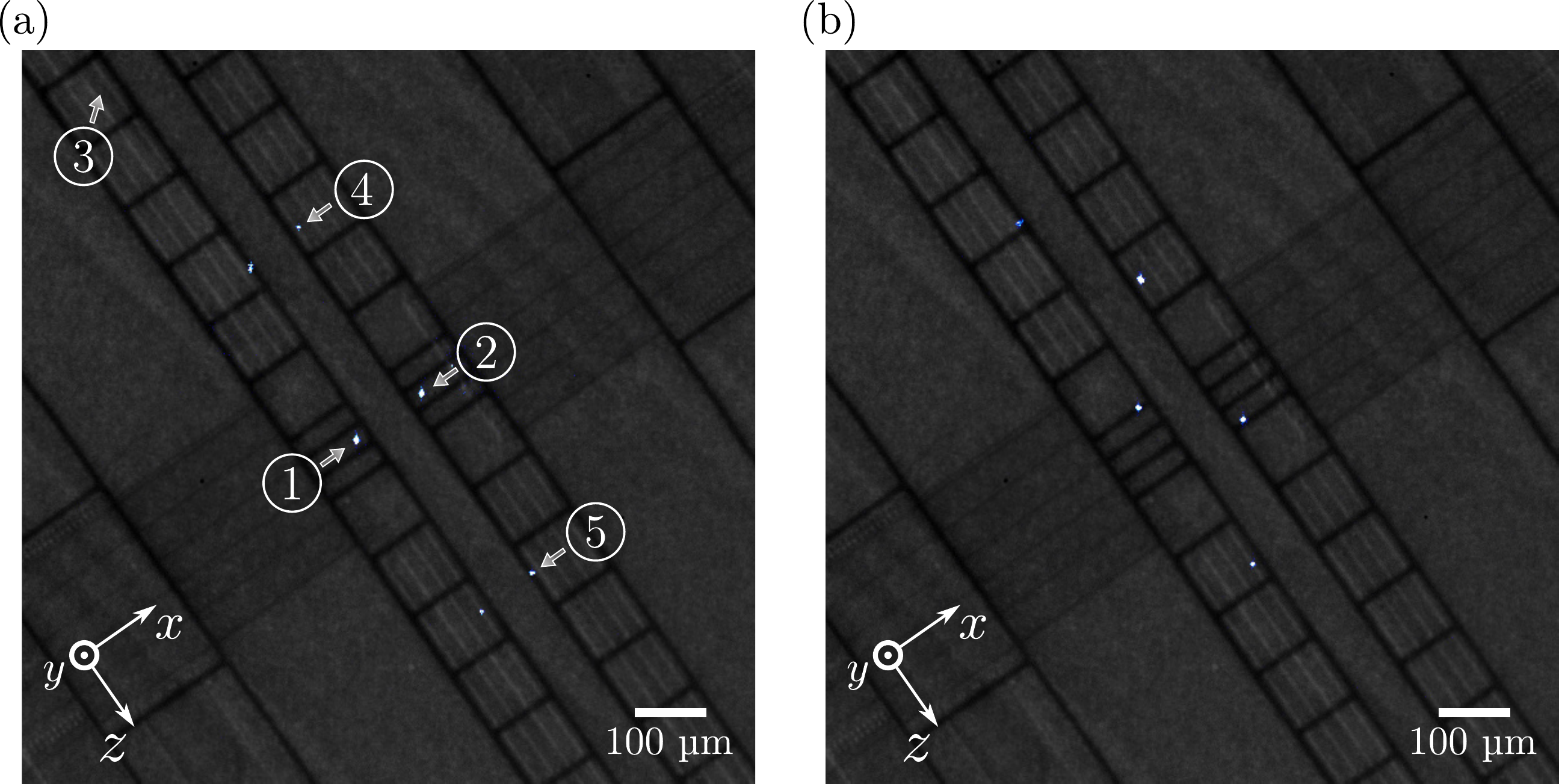}
	\caption{Images of ions simultaneously trapped in different lattice configurations. The images of ion fluorescence at \SI{397}{\nano\meter} (blue-white) is superimposed onto a background image of the trap electrodes (gray scale), obtained by illumination with a $\SI{395}{\nano\meter}$ LED source. (a) Trapping of 6 ions in a rectangular lattice configuration. The axial ion spacing in the left and right multiwells is $\trapdistz \approx \SI{300}{\micro\metre}$, and the spacing between the two multiwells is $\trapdistx \approx \SI{100}{\micro\metre}$. (b) Trapping of 5 ions in a triangular lattice configuration. The ion spacing, nominally identical to (a), is not perfectly uniform due to stray electric fields.}
	\label{meas:ion-images}	
\end{figure}

In a first experiment, we investigate the ability of the twin-trap to confine ions in different lattice configurations. Figure~\ref{meas:ion-images} shows images of ions, simultaneously trapped in multiple trapping sites. The ion-surface separation is $d=\SI{120}{\micro\meter}$. In figure~\ref{meas:ion-images}\,(a), 6 ions are trapped in a rectangular lattice with trap spacings $s_x\approx\SI{100}{\micro\meter}$ and $s_z\approx\SI{300}{\micro\meter}$. Ions at the center (sites 1 and 2) are brighter than the ions further out mainly due to a small tilt of the major axis of the elliptical imaging beam relative to the trap surface and partly due to different micromotion conditions. Figure~\ref{meas:ion-images}\,(b) shows 5 ions trapped in a triangular lattice configuration which results from the rectangular lattice in (a) by a shift of the left and right DC multiwells by a quarter lattice period in opposite directions. To trap ions in multiple lattice sites, we employ a combination of two loading techniques. First, the two photoionization beams at \SI{422}{\nano\meter} and \SI{379}{\nano\meter} are sequentially directed to the trapping sites where single ions are to be trapped. Loading ions in some of the trapping sites was difficult, which we attribute to stray electric fields. These sites were filled using shuttling of ions from adjacent sites. 

We further demonstrate shuttling of an entire ion lattice: In the video in the Supplemental Material, we show a simultaneous translation of a rectangular lattice of 4 ions over a distance of one lattice period, $s_z=\SI{306}{\micro\meter}$. While all ions remain trapped during the transport, three of the four crystallized ions temporarily melt. This mainly happens due to an asynchronous update of the different DC voltages provided by the supply; in parts also due to a variation of the stray electric field. We emphasize that during the shuttling we applied only a constant, global micromotion compensation field. The successful transport thus indicates a relatively constant stray electric field over the entire shuttling distance of $\SI{600}{\micro\meter}$. A total of only 8 DC control voltages is used for the shuttling process. The low shuttling speed of about \SI{6}{\micro\meter\per\second} is limited by the slew rate of the stable DC voltage supply used to drive the electrodes. 

After several weeks of trap operation, we observed a substantial change of the stray electric field, resulting in ion loss during shuttling operations. Using a single ion as a probe we characterized the spatial variation of the stray electric field at sites 4, 2 and 5 (cf. figure~\ref{meas:ion-images}\,(a)). The measurement is done by adjusting the micromotion compensation voltages to maximize ion fluorescence at \SI{397}{\nano\meter} close to the atomic transition frequency. The stray field is then given by the compensation field, with opposite sign. The data, listed in table~\ref{tab:results:stray fields}, reveal that the stray field component $E_x$ in the central site 2 has a five times larger amplitude than in sites 4 and 5 and is pointing in opposite direction. The component $E_y$ is significantly smaller than $E_x$ and approximately constant for all sites. The precision of the measurement of $E_y$ is lower than that of $E_x$. The \SI{397}{\nano\meter} beam used to detect stray-field induced micromotion propagates in the $xz$-plane, parallel to the trap surface, and is not sensitive to micromotion in the $y$-direction. One axis of the RF quadrupole field is tilted by only $\alpha\approx\ang{22}$ from the $y$-direction.

In addition, ions could be loaded in sites 1 and 2 at the chip center without applying axial confinement. The residual axial frequency $\wz\approx2\pi\times\SI{600}{\kilo\hertz}$, independent of the applied RF voltage, stayed approximately constant over the whole trap operation period. As zones 1 and 2 where often used for ion loading, this stray confining field may have been caused by laser-induced charges or by inhomogeneous contamination arising from the loading process \cite{Har2010,Nar2011,Bra2012,War2013}.

Finally, heating rate measurements were performed to further explore the potential of the linear twin-trap prototype for ion-ion coupling. The measurements were taken in sites 1, 2 and 3 (cf. figure~\ref{meas:ion-images}\,(a)) using the sideband-ratio method \cite{Lei03}. The results are listed in table~\ref{tab:results:heating-rates}. The measured values, obtained at axial frequencies $\wz\approx2\pi\times\text{(1.2 - 1.5)}\,\si{\mega\hertz}$, are in a range $\Gh\approx\text{(100 - 500)}\,\si{phonons/\second}$ for the three trapping sites. Given the targeted ion-ion coupling rate $\Wc\approx2\pi\times\SI{1}{\kilo\hertz}$, these heating rates should allow the observation of ion-ion coupling on a few quanta level \cite{Bro2011,Har2011}. However, to harness the coupling for spin-spin interactions or high-fidelity entangling operations between ions in adjacent sites a significantly lower heating rate would be required. A further characterization is necessary to determine whether the measured heating rates are limited by technical noise that could be filtered out or by surface noise. In fact, surface contamination is a possible reason for the high heating rates. While the trap chip has been cleaned of photoresist residues and dicing debris at the Infineon facilities, no further cleaning steps were done prior to loading into the vacuum chamber. Additional chemical cleaning or ex situ surface treatments \cite{McK2014,Sed2018} could lower the observed heating rates. Also, a change in electrode material from AlSiCu to a noble metal might significantly reduce the experienced heating due to the absence of native oxide layers \cite{Kum2016-2}. We currently work on a new chip version with gold electrodes. Another option would be in situ cleaning of trap electrodes by argon ion bombardment which has been reported to drastically lower the heating rate \cite{Hit2012,Dan2014}.

\begin{table}[h]
	\parbox{.45\linewidth}
	{
		\centering
		\caption{Stray electric field components $E_x, E_y$ at different trapping sites.}	
		\label{tab:results:stray fields}
		\begin{threeparttable}
			\begin{tabular}{cccc}
				\toprule
				site & $E_x$ (\si{\volt\per\meter}) & $E_y$ (\si{\volt\per\meter})\\
				\midrule\midrule
				4 & \num{+174\pm15} & $\approx60$ \\
				\midrule
				2 & \num{-640\pm30} & $\approx60$ \\
				\midrule
				5 & \num{+116\pm10} & $\approx60$ \\		
				\bottomrule
			\end{tabular}
		\end{threeparttable}
	}
	\hfill
	\parbox{.4\linewidth}{
		
		\centering
		\caption{Ion heating rate at different trapping sites.}
		\label{tab:results:heating-rates}	
		\begin{threeparttable}
			\begin{tabular}{ccc}
				\toprule
				site & $\wz/(2\pi)$ & \Gh (\si{phonons\per\second})\\
				\midrule\midrule
				1 & \SI{1.45}{\mega\hertz} & \num{288\pm 35} \\
				\midrule
				2 & \SI{1.48}{\mega\hertz} & \num{472\pm 50} \\
				\midrule
				3 & \SI{1.24}{\mega\hertz} & \num{131\pm 13} \\							
				\bottomrule
			\end{tabular}
		\end{threeparttable}
	}
\end{table}


\section{Conclusion}
\label{sec:conclusion}

In summary, we have proposed, built and operated a new design of an ion-lattice quantum processor based on two-dimensional arrays of linear surface traps. A core aspect of our approach is the usage of ion-shuttling operations in two spatial dimensions that enable a dynamical configuration of the ion lattice in terms of lattice connectivity and ion-spacing. The latter enables tunable interactions between ions in adjacent lattice sites. We have shown the feasibility of our approach by means of detailed trap simulations of a simplest-instance version, consisting of two parallel linear traps with $2\times9$ trapping sites. The simulated trapping potentials facilitate interaction strengths between ions in adjacent sites in the \si{\kilo\hertz} range, while maintaining a moderate ion-surface separation $d=\SI{120}{\micro\meter}$ to keep the electric field noise low. We demonstrate the scalability of this design with additional simulations of an array with $10\times10$ sites, shown in appendix~\ref{app:large_array}. We have built several versions of the $2\times9$ array in an industrial facility using multilayer microfabrication. The employed fabrication processes are compatible with further scaling-up the array size where the growing number of island-like electrodes will require a more dense routing: up to 6 metal layers can readily be realized, and even more layers are possible by adding planarization steps. In the future, our CMOS fabrication process could also be extended to include waveguide structures for integrated optical addressing of single ions and pairs of ions \cite{Meh2016}.

We have experimentally demonstrated the basic operability of a prototype device with $2\times9$ trapping sites, showing simultaneous trapping of ions in multiple lattice sites, DC voltage-controlled shuttling and resolved-sideband operations (heating rate measurements). The cooling beams were elliptical to cover multiple trapping sites at once; in the future steerable beams or multiple beams \cite{Buh2003} may be employed to reduce the optical power needed. We have further demonstrated the ability to configure the ion lattice, showing trapping in a rectangular lattice and a triangular lattice configuration and translation of an entire ion lattice by one lattice period. This configurability is only possible in a linear trap array and is one of the principle points of our design. For shuttling along the trap axis we have employed a periodic voltage assignment to the trap's DC segments, which allows one in principle to axially transport ion sub-lattices over arbitrary distances using only a small number of DC control voltages. The shuttling speed, being currently limited by the stable DC supply, could in the future be increased by orders of magnitude using a faster supply \cite{Row2002,Bow2012,Wal2012}. Axial translations can also be employed as a technique for fast sequential loading of an entire ion lattice: ion loading takes place at one dedicated site per linear trap and loaded ions are subsequently shuttled together with all other ions in the multiwell to the adjacent site using axial translations (cf. section~\ref{sec:twin-trap:simulation:translations}). This technique does not require ionization beams to be steered across the array and could be combined with a pre-cooled source of atoms to further increase the loading rate \cite{Bru2016}. A draw-back of the periodic voltage assignment is the limited control of the trapping potential at different lattice sites. Lattice translations in our prototype design using only a global micromotion compensation field were successful at first, but were eventually limited by a spatially-varying stray electric field. Indeed, we find the vulnerability to stray charges to be the biggest limitation of our prototype device. This problem can be tackled in future chip versions: First, the creation of stray charges on exposed dielectrics can be inhibited by reducing the electrode gap size (currently \SI{9}{\micro\meter}) and by using a noble metal for the top metal layer, \eg gold. Second, the electrode design can be adapted to allow for a larger number of control electrodes for independent micromotion compensation in more lattice sites.
Another limitation of our prototype device is the relatively high heating rate $\Gh\sim\text{(100-500)}\,\si{phonons\per\second}$ at $\wz\approx2\pi\times\SI{1.5}{\mega\hertz}$, which is only slightly smaller than the targeted ion-ion coupling rate $\Wc\sim2\pi\times\SI{1}{\kilo\hertz}$. Such a heating rate does not allow for the ion-ion coupling to be used for quantum simulations. We emphasize that the heating rate in our setup is not limited by Johnson noise from the trap electrodes as the electric field noise estimates based on the resistance measurements show. We have discussed several means to reduce the heating rate, particularly by changing the electrode material and by applying surface cleaning procedures.

\section*{Acknowledgements}

This project has received funding from the European Union’s Horizon 2020 research and innovation programme under Grant Agreement No.~801285 (PIEDMONS). We further acknowledge financial support by the Austrian Science Fund (FWF) through projects P26401 (Q-SAIL) and F4016-N23 (SFB FoQuS), and by the Institut f\"ur Quanteninformation GmbH.

\section*{Appendices}
\appendix

\section{Calculation of RF and DC voltages}
\label{app:voltage-sets}

The twin trap's RF drive parameters are chosen in the following way. First, a maximally applicable RF voltage $\Urf\approx\SI{400}{\volt}$ is assumed. In the configuration with reduced RF voltage
on the inner RF rail, figure~\ref{fig:sim_confinement_interaction}\,(a), where the trap efficiency is decreased, the drive frequency \Wrf is then set to yield a stability factor of $q = 0.4$. Keeping \Wrf constant, the RF
voltage \Urf is then adjusted to achieve $q = 0.4$ in the default configuration, figure~\ref{fig:sim_confinement_home}.

\begin{figure}[hb]
	\centering
	\includegraphics[width=0.45\textwidth]{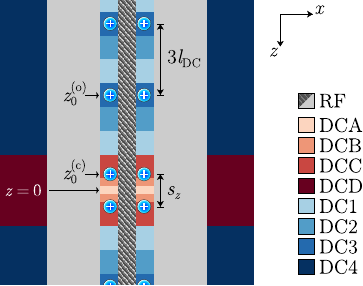}
	\caption{Grouping of electrode segments for the calculation of voltage sets in the axial interaction zone.}
	\label{fig_app:electrodes_axial-interaction}	
\end{figure}

For the simulation of DC multiwell confinement and ion shuttling, we use an algorithm that calculates DC voltage sets for axial confinement and micromotion compensation simultaneously at two arbitrary trapping positions $\bm{r}_0^\text{(l)}$ and $\bm{r}_0^\text{(r)}$ in the left and right RF null, respectively. This includes different axial trapping positions $z_0^\text{(l)}\neq z_0^\text{(r)}$. The voltage set for confinement at these two sites automatically creates additional sites with a spacing of $3\,\lDC$ along the trap axes due to the periodic assignment of voltages to the DC segments. Necessary conditions for a trapping site at position $\bm{r}_0$ are a vanishing axial electric field, $E_z(\bm{r}_0)=0$, and a positive curvature, $\partial_z^2\phi(\bm{r}_0)>0$. In addition, $\bm{r}_0$ needs to be overlapped with the RF null, \ie $E_{x,y}(\bm{r}_0)=0$. The sets for micromotion compensation require control over the radial electric field components $E_{x,y}(\bm{r}_0)$. A shift of the trapping position along $z$ can be realized by the axial field component $E_{z}(\bm{r}_0)$. This amounts to 8 field parameters (6 electric field components and 2 curvatures) for the two trapping sites at $\bm{r}_0^\text{(l)}$ and $\bm{r}_0^\text{(r)}$. Let now $\bm{b}$ be a vector of the desired 8 field parameters. Further, let $\bm{x}$ be the unknown vector of voltages applied to the set of DC electrodes that produces $\bm{b}$. Then it holds $\bm{b} = A \bm{x}$, where the entries in the square matrix $A$ are the contributions of the individual electrodes to the 8 field parameters. These entries are determined by trap simulation. The unknown voltage set $\bm{x}$ is then found by inversion of matrix $A$. This method only succeeds if $A$ is of full rank, which requires at least 8 electrodes whose field and curvature contributions are linearly independent. For the simulations of trap confinement in the default configuration, figure~\ref{fig:sim_confinement_home}, and for the simulation of independent axial translations, figure~\ref{fig:sim_characterization_BBshuttling}, the DC segments are grouped in 8 independent electrodes, as shown in figure~\ref{fig:sim_characterization_BBshuttling}\,(a). For the simulation of confinement in the axial interaction zone, figure~\ref{fig:sim_confinement_interaction}, the segments are differently grouped, as shown in figure~\ref{fig_app:electrodes_axial-interaction}.
 Here, the voltages on the electrodes are assumed to have a mirror symmetry along both the central RF rail and along the $x$-axis through the center of the trap. In this way, the control of two trapping sites, one at $z_0^\text{(c)}$ within the interaction zone and the other at $z_0^\text{(o)}$ in the outer region, is sufficient to create multiwells across the entire length of the trap. The position $z_0^\text{(c)}$ of the first site sets the reduced axial distance $s_z$ between the central trapping sites. The position $z_0^\text{(o)}$ of the second site controls the location of all the outer trapping sites, which have a fixed spacing given by the trap spacing $3\lDC$.

\section{Calculation of Johnson noise and RF pickup}
\label{app:field-noise}

In this section, the estimates for the ion heating rate due to Johnson noise in the trap electrodes, as well as the magnitude of the RF pickup voltage on the DC electrodes are derived. For the estimate of the heating rate we consider the leads for the periodically connected island electrodes on the metal 2 layer, which have by far the largest resistance on the trap chip. These leads have a maximal length between bonding pad and furthest DC segment of about $l=\SI{3.56}{\milli\meter}$ and a
width of $w=\SI{20}{\micro\meter}$, resulting in a resistance of $R_\mathrm{lead}= l\rho_\textrm{metal\,2}/(w t)\approx \SI{0.46}{\ohm}$ at $T \approx \SI{20}{\kelvin}$, where $t=\SI{1000}{\nano\meter}$ is the thickness of the metal 2 layer. The via resistances can be neglected. The amount of electric field noise created by this resistance at the position of a trapped ion is \cite{Bro2015} $S_E^\textrm{(JN)}=4k_\textrm{B}TR_\mathrm{lead}/\delta_\textrm{c}^2=\SI{1.06e-16}{\volt^2\meter^{-2}\hertz^{-1}}$, where $k_\textrm{B}$ is the Boltzmann constant and $T=\SI{20}{\kelvin}$. The characteristic distance of the segmented DC electrode, $\delta_\textrm{c}$, is found by trap simulation and has a maximal value $\delta_\textrm{c}=\SI{2.19}{\milli\meter}$ along the axial direction for all axial positions (for the radial directions, $\delta_\textrm{c}$ is at most about a factor 2 smaller). This electric field noise corresponds to an axial heating rate of \cite{Bro2015} $\Gh^\textrm{(JN)}=Q^2 S_E^\textrm{(JN)}/(4M\hbar\wz)\approx \SI{0.015}{phonons\per\second}$, where $Q$ and $M$ and the charge and mass of a \Ca ion, $\hbar$ is the reduced Planck constant and $\wz=2\pi\times\SI{1}{\mega\hertz}$ is the ion's axial frequency. 

\begin{figure}[htbp]
	\centering
	\includegraphics[width=0.47\textwidth]{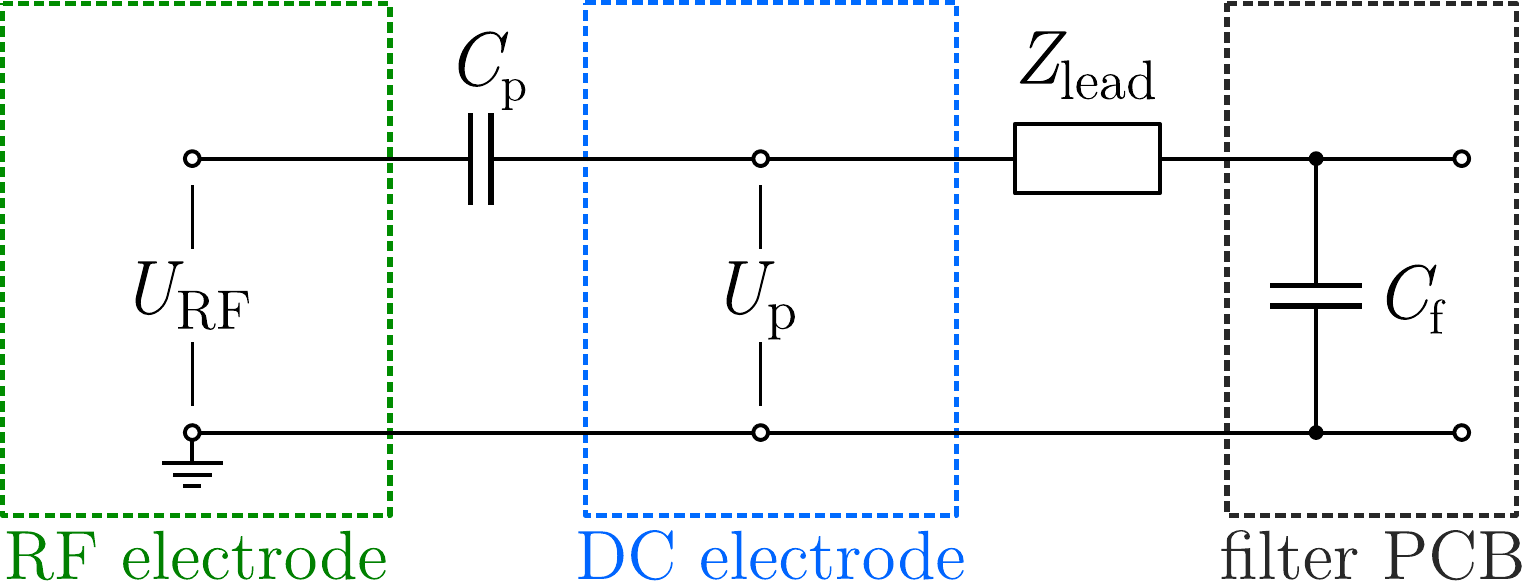}
	\caption{Model circuit for the RF grounding of a DC trap electrode. The parasitic capacitance $C_\textrm{p}$ leads to an RF pickup voltage $U_\textrm{p}$ at the DC electrode. The magnitude and phase of $U_\textrm{p}$ depend on the connection line impedance $Z_\textrm{lead}$ and on the filter capacitance $C_\textrm{f}$.}
	\label{fig_app:RF-pickup}	
\end{figure}

For the estimate of the RF pickup voltage on the trap's DC electrodes we consider the simplified electrical circuit in figure~\ref{fig_app:RF-pickup}.
The RF drive voltage \Urf is applied to the trap's RF
electrode (green box). The parasitic capacitance $C_\textrm{p}$ between the trap electrodes couples
the DC electrode (blue box) to the RF electrode, leading to an RF pickup voltage $U_\textrm{p}$
on the DC electrode. The value of $U_\textrm{p}$ depends on how well the DC electrode is connected to GND, $U_\textrm{p}=\epsilon_\textrm{p}\Urf$, with the complex RF pickup ratio 
\begin{equation}
\label{eq:app_pickup-ratio}
\epsilon_\textrm{p} = \frac{Z_\textrm{lead} + Z_{C_\textrm{f}}}{Z_{C_\textrm{p}} + Z_\textrm{lead} + Z_{C_\textrm{f}}}\,,
\end{equation}
and $Z_C = -i/(\Wrf C)$ being the impedance of a capacitance $C$ at frequency \Wrf. To give an upper bound on the pick up ratio $\epsilon_\textrm{p}$, we consider one of the periodically connected island electrodes which have the largest parasitic coupling capacitance  $C_\textrm{p}$ and largest lead impedance $Z_\textrm{lead}$. We estimate the parasitic coupling capacitance $C_\textrm{p}\approx\SI{0.01}{\pico\farad}$ from finite element simulations of the trap geometry\footnote{Finite element simulations were performed with COMSOL Multiphysics, Version 5.3a.} (cf. figure~\ref{fig:fab_traps}\,(d)). The lead inductance $L_\textrm{lead}\approx\SI{0.2}{\nano\henry}$ is calculated from the simulated capacitance matrix \cite{Pau1976}. The lead resistance, calculated above, is $R_\mathrm{lead}\approx \SI{0.46}{\ohm}$ and dominates the lead impedance $Z_\textrm{lead}=R_\mathrm{lead} + i\Wrf L_\textrm{lead} \approx \SI{0.46+i 0.03}{\ohm}$ at the RF drive frequency $\Wrf=2\pi\times\SI{25}{\mega\hertz}$. The grounding capacitance $C_\textrm{f}\approx\SI{330}{\nano\farad}$\footnote{Kemet, C2220C334J1GACTU} is given by the capacitance of the low-pass filters used in our setup. These filters are located on a printed circuit board (PCB) within the cryogenic setup, only a few \si{\centi\meter} from the trap chip. Finally, assuming that the connection line impedance is dominated by the lead impedance $Z_\textrm{lead}$, we arrive at an upper bound for the RF pickup ratio of $|\epsilon_\textrm{p}|\approx\num{7.2e-7}$.

\section{Simulation of a linear trap array with 10$\,\times\,$10 trapping sites}
\label{app:large_array}

In this section, we show that the twin-trap design, figure~\ref{fig:sim_electrode_geometry}, can be extended to a larger number of parallel linear traps. For this, multiwell confinement and RF shuttling in a linear trap array with $10\times10$ trapping sites are simulated. DC shuttling along the axial direction is not simulated since this aspect is already covered by the studies in the twin-trap: confinement with reduced axial distance, figure~\ref{fig:sim_confinement_interaction}\,(c), and independent axial translations of two adjacent DC multiwells with 9 trapping sites each, figure~\ref{fig:sim_characterization_BBshuttling}. It should be emphasized, that the simulations presented here are intended only as a proof-of-principle study. The electrode geometry is not optimized and can be further improved.

\begin{figure}[htbp]
	\centering
	\includegraphics[width=0.9\textwidth]{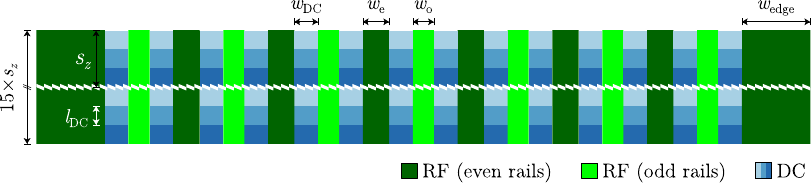}
	\caption{Electrode geometry of the a linear trap array with 100 trapping sites. The width of the RF rails is $w_\text{e}=\SI{88}{\micro\meter}$ and  $w_\text{o}=\SI{70.4}{\micro\meter}$ for the even and odd rails, respectively. The outermost RF rails have an increased width of $w_\text{edge}=\SI{228.8}{\micro\meter}$. The DC rails have a width of $w_\text{DC}=\SI{79.2}{\micro\meter}$. The DC segment length of $l_\text{DC}=\SI{74.8}{\micro\meter}$ gives rise to a DC multiwell periodicity of about $\trapdistz=\SI{224}{\micro\meter}$. An additional ground electrode parallel to the trap surface (not shown) is located at a vertical distance of $y=\SI{1.0}{\milli\meter}$.}
	\label{fig_app:geometry_big-array}	
\end{figure}

The geometry of the simulated $10\times10$ trap array, is shown in figure~\ref{fig_app:geometry_big-array}. RF confinement in the radial ($xy$-) plane is produced by parallel RF rails with alternating widths $w_\text{e}=\SI{88}{\micro\meter}$ and $w_\text{o}=\SI{70.4}{\micro\meter}$, referred to as even and odd RF rails, respectively, in what follows. A total of 15 RF rails leads to 14 parallel linear traps, out of which the innermost 10 linear traps are used for ion storage. The widths of the even and odd RF rails differ by about 20\%. This leads to a tilt of the radial modes with respect to the trap normal in the presence of DC confinement, allowing for simultaneous laser cooling of all secular modes with laser beams parallel to the trap surface. The segmented DC rails have a width $w_\text{DC}=\SI{79.2}{\micro\meter}$ and a segment length $l_\text{DC}=\SI{74.8}{\micro\meter}$. Like in the twin-trap design, the segments are periodically connected, with the same voltage being applied to every third segment. This allows one to create DC multiwell confinement with a well period of $3\lDC\approx\SI{224}{\micro\meter}$.

Inhomogeneities of the RF potential across the array caused by edge effects are mitigated in three ways: First, an additional GND electrode at a distance $y=\SI{1.0}{\milli\meter}$ above the trap surface is introduced, which also increases the trap depth by roughly a factor 1.5, compared to a design without top GND layer. A top GND electrode could be realized for instance with a glass plate coated with indium tin oxide (ITO) and mounted rigidly above the trap chip. ITO remains conductive and optically transparent at cryogenic temperatures \cite{Wie2017}. Second, an additional pair of linear dummy traps is added at either side of the array. The 10 central linear traps used for the quantum register are thereby increased to 14 linear traps. Ions loaded accidentally in the outer dummy traps could be deterministically pushed out by using, for instance, suitable DC control fields on the outermost DC electrodes. Third, the width of the outermost RF rails is increased to $w_\text{edge}=\SI{228.8}{\micro\meter}$. We note that the simplified geometry in figure~\ref{fig_app:geometry_big-array} only shows the minimum of DC electrodes necessary for creating a 2\,D ion lattice. For a realistic operation as ion-lattice quantum processor, a further segmentation of the DC rails would be necessary. In particular, control electrodes for stray electric field compensation (micromotion compensation) and for fine control of secular frequencies (and potentially mode orientations) would be required.

\subsection*{Multiwell confinement}
\begin{figure}[htbp]
	\centering
	\includegraphics[width=\textwidth]{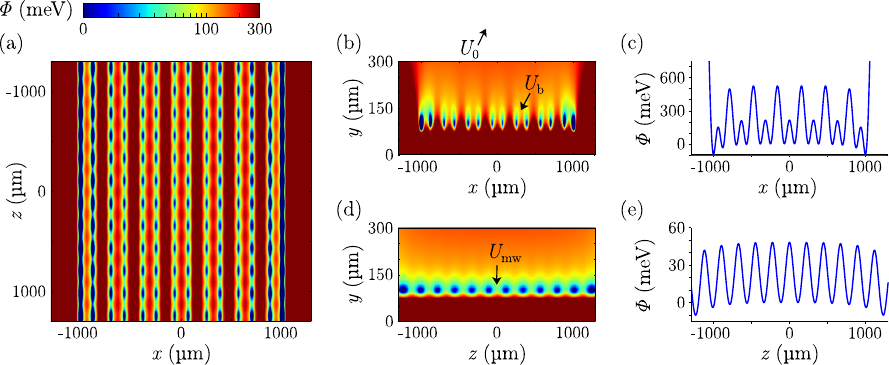}
	\caption{Trap confinement in the default configuration of the array with $10\times10$ trapping sites. Subplots (a), (b), (d) show cross sections of the total potential $\mathit{\Phi}$ in the $xz$-, $xy$- and $zy$-plane, respectively, crossing the trapping site at $r_0 = (-67, 102, -112)\,\si{\micro\meter}$. The color scale is cut off at \SI{300}{\milli\electronvolt}, non-linear for better visibility of the minima and we set $\mathit{\Phi}(r_0)=0$. (c) Potential along the $x$-direction through the central trapping site at $x_0, z_0=(-67,-112)\,\si{\micro\meter}$. (e) Axial multiwell potential through the same site.}
	\label{fig_app:sim_big-array}	
\end{figure}

To simulate multiwell confinement, a voltage set for axial confinement is calculated for a single trapping site at the center of the array. Upon applying this set, the periodicity of the RF and DC electrodes automatically creates a rectangular array of trapping sites. DC voltages are applied to the DC segments as well as to the RF rails, in order to gain the required number of control parameters for axial confinement and micromotion compensation. In the default trapping configuration, an equal RF voltage \Urf is applied to the even and odd RF rails and a DC voltage set for axial multiwell confinement is applied. The total confining potential $\mathit{\Phi}$ in this configuration is shown in figure~\ref{fig_app:sim_big-array}. The cross sections (a), (b), (d) show a rectangular lattice of $14\times12$ trapping sites out of which the central $10\times10$ sites are to be used for ion storage. The additional sites at the trap edges are dummy sites. The ion-surface separation of the central $10\times10$ sites is about $d\approx\SI{102}{\micro\meter}$. An RF voltage of $\Urf=\SI{172}{\volt}$ at $\Wrf=2\pi\times\SI{30}{\mega\hertz}$ yields a stability $q$-factor of 0.4. The DC voltages for axial confinement are on the order of \SI{1}{\volt}. The secular frequencies are $\wz=2\pi\times\SI{1.0}{\mega\hertz}$ axially and $\wrado, \wradt=2\pi\times(4.0,4.4)\,\si{\mega\hertz}$ radially, with a radial mode tilt $\thetar=\ang{8.4}$ with respect to the surface normal. The axial mode is aligned with the $z$-axis, $\thetaz=0$. The axial multiwell barrier $\Umw=\SI{45}{\milli\electronvolt}$, the RF barrier $\Ub=\SI{116}{\milli\electronvolt}$ and the global trap depth $U_0=\SI{330}{\milli\electronvolt}$ all have high values, well above the average kinetic energy $E_\text{th}\approx\SI{26}{\milli\electronvolt}$ of thermal gas molecules at room-temperature. The global trap depth $U_0$ is defined as the potential $\mathit{\Phi}$ at the position of the top GND layer, $y=\SI{1}{\milli\meter}$.
The inner $10\times10$ trapping sites show a very good homogeneity: The variation in ion-surface separation $d$ is about \SI{1}{\micro\meter}. The stability $q$-factor varies within 0.404 and 0.409 for all sites. Variations in secular frequencies are about \SI{4}{\kilo\hertz} axially and \SI{30}{\kilo\hertz} radially. The radial mode tilt varies within \ang{8.0} and \ang{13.1}, the axial mode tilt stays below \ang{0.1}. Radial shift of the sites off the RF null are below \SI{1}{\micro\meter}. Variations in the trap depths are negligible. 

\subsection*{RF shuttling}
\begin{figure}[htbp]
	\centering
	\includegraphics[width=0.86\textwidth]{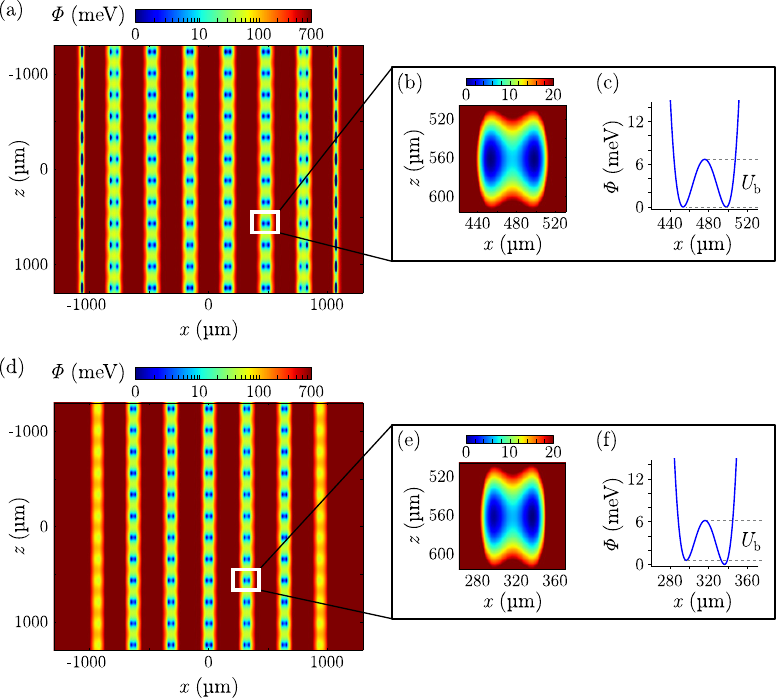}
	\caption{Trap confinement in the reduced RF configuration with $s_z\approx\SI{40}{\micro\meter}$, achieved by attenuating the RF voltage on the even RF rails, (a)-(c), and on the odd rails, (d)-(f). Panels (a) and (d) show cross sections of the total potential $\mathit{\Phi}$ in the $xz$-plane, crossing the trapping site at $\bm{r}_0=(139,98,-112)\,\si{\micro\meter}$ and $\bm{r}_0=(20,101,-112)\,\si{\micro\meter}$, respectively. The color scale is non-linear, a quartic potential has been subtracted from the data for better visibility of the minima and we set $\mathit{\Phi}(r_0)=0$. The insets show a magnified view, (b) and (e), of the marked pair of trapping sites, and the line potential through them, (c) and (f).}
	\label{fig_app:sim_big-array_shuttling}	
\end{figure}

Entanglement between ions in adjacent linear traps is facilitated by a reduction of the distance $s_x$ between adjacent RF nulls. This is achieved by reducing the RF voltage \Urf on either the even or the odd RF rails. At a separation $s_x=\SI{40}{\micro\meter}$ one calculates a motional coupling rate $\Wc=2\pi\times\SI{1.4}{\kilo\hertz}$, using equation~(\ref{eq:coupling-rate}) and assuming an axial frequency $\wz=2\pi\times\SI{1}{\mega\hertz}$. 
Once the reduced distance $s_x$ is reached, the axial mode frequencies of ions that are to be coupled are tuned into resonance using DC control fields; unwanted coupling, e.g. between non-nearest neighbors is avoided by detuning the frequencies of these wells \cite{Bro2011,Har2011,Wil2014}, as outlined in appendix~\ref{app:parasitic_coupling}. The trap confinement at the reduced distance $s_x$ is shown in figure~\ref{fig_app:sim_big-array_shuttling}.
The cross sections (a) and (d) show how the $14\times12$ trapping sites are rearranged upon attenuation of the RF voltage on the even and odd RF rails, respectively. In both configurations, the sites form pairs of columns such that for any trapping site a reduced distance $s_x\approx\SI{40}{\micro\meter}$ to either the adjacent site on the right or on the left can be realized\footnote{In figures~\,\ref{fig_app:sim_big-array_shuttling}\,(a) and (d), a quartic potential $\Phi_\text{offset}(x,z) = a_x x^4 + b_x x^2 + a_z z^4+\text{const.}$ is subtracted from the data to increase the visibility of the minima. For the RF reduction on the even RF rails, $a_x=\SI{6.0e-15}{\electronvolt\per\micro\meter^4}$, $b_x=\SI{2.0e-9}{\electronvolt\per\micro\meter^2}$, $a_z=\SI{4.5e-15}{\electronvolt\per\micro\meter^4}$. For the RF reduction on the odd RF rails, $a_x=\SI{4.0e-14}{\electronvolt\per\micro\meter^4}$, $b_x=\SI{1.5e-8}{\electronvolt\per\micro\meter^2}$, $a_z=\SI{4.5e-15}{\electronvolt\per\micro\meter^4}$.}. The ion-surface separation is in both configurations about $d\approx\SI{100}{\micro\meter}$, almost identical to the default configuration in figure~\ref{fig_app:sim_big-array}. In general, the ion-surface separation is practically unchanged during RF shuttling. The insets, figure~\ref{fig_app:sim_big-array_shuttling}\,(b) and (e) show a magnified view of the marked pairs of trapping sites. The double-well potentials connecting the two sites of each pair are shown in (c) and (f). In the two configurations, the RF voltage is either attenuated by about 59.4\% on the even RF rails, or by 41.4\% on the odd rails, relative to the respective other rail which is at $\Urf=\SI{350}{\volt}$. The difference in required RF attenuation for the two configurations stems from the different RF rail widths. In either configuration, the axial multiwell confinement can be maintained using DC voltages on the order of \SI{1}{\volt} with standard secular frequencies of $\wz=2\pi\times\SI{1.0}{\mega\hertz}$ axially and $\wrad=2\pi\times\text{(2.0 - 3.0)}\,\si{\mega\hertz}$ radially. The axial mode remains aligned with the $z$-axis, $\thetaz=0$, the radial mode tilt is increased to about $\thetar\sim\ang{35}$. The reason for the smaller radial frequencies in comparison to the default configuration is the decreased trap efficiency, just as in the case of the twin-traps. For the simulations, a maximally applicable RF voltage $\Urf=\SI{350}{\volt}$ was assumed, limiting the stability $q$-values to 0.21 and 0.26, respectively. Likewise, the double-well barrier is limited to $\Ub\approx\SI{3.9}{\milli\electronvolt}$ and $\Ub\approx\SI{6.1}{\milli\electronvolt}$. The multiwell barrier $\Umw\approx\SI{35}{\milli\electronvolt}$ and the trap depth $U_0=\text{(192, 334)}\si{\milli\electronvolt}$ remain at large values. 

Concerning the homogeneity of the central $10\times10$ trapping sites, the most notable variation exists in the reduced trap spacing $s_x$. For attenuation on the even RF rails, this distance changes from $s_x=\SI{39}{\micro\meter}$ at the central linear traps, $x\approx\SI{\pm158}{\micro\meter}$, to $s_x=\SI{44}{\micro\meter}$ at the next pair of traps, $x\approx\SI{\pm476}{\micro\meter}$. For attenuation on the odd RF rails, the distance changes from $s_x=\SI{40}{\micro\meter}$ at the array center, $x=\SI{0}{\micro\meter}$, to $s_x=\SI{39}{\micro\meter}$ at the next pair of traps, $x\approx\SI{\pm316}{\micro\meter}$, and further to $s_x=\SI{35}{\micro\meter}$ at the outer pair of traps, $x\approx\SI{\pm632}{\micro\meter}$. The variation in trap spacing $s_x$ is caused by the edge effects of the trap array along the $x$-direction and limits the possibility of parallelized entangling operations across the entire lattice due to the difference in expected coupling rate \Wc. Edge effects due to the finite number of RF rails also cause a variation in the stability $q$-factor and, therefore, in the radial frequencies. For attenuation on the even 
rails, the $q$ values are between 0.21 and 0.35, allowing for simultaneous stable trapping, and the radial frequencies vary within $\wrad=2\pi\times\text{(2.0 - 3.6)}\,\si{\mega\hertz}$. For attenuation on the odd rails, the effect is weaker, with the $q$-factor ranging between 0.26 and 0.22 and a radial frequency 
variation of $\wrad=2\pi\times\text{(2.2 - 2.8)}\,\si{\mega\hertz}$. The differences in \wrad and in trap spacing $s_x$ across the array 
cause a variation in the double-well barrier \Ub. However, \Ub does not fall below \SI{2}{\milli\electronvolt} across the entire array and in both configurations. For motional coupling between adjacent sites, the variation in radial frequencies is not a concern if the axial mode is employed. The finiteness of the array leads to additional inhomogeneities in conjunction with the fact that the voltage set for axial confinement is calculated only for a single site at the array center. This makes the axial multiwell potential non-ideal at the array edges. Due to this, the axial frequency \wz varies by \SI{74}{\kilo\hertz} for attenuation on the even RF rails. For attenuation on the odd rails, the effect is significantly smaller with a variation in \wz of \SI{7}{\kilo\hertz}. Lastly, edge effects lead to small shifts of the trapping sites off the RF null of about \SI{1}{\micro\meter} for both configurations, comparable to the default configuration.

We note that many of the above mentioned limitations could be mitigated in an optimized trap geometry. In particular in the reduced RF configuration, one could achieve a much better homogeneity of the reduced trap spacing $s_x$ across the array, and therefore of the ion coupling strength \Wc, by adjusting the RF and DC rail widths. A first step in this direction was made by increasing the width of the outermost RF rails, allowing for a match of the stability $q$-values of the linear traps in the default configuration. Even with remaining variations in \Wc across the array, parallelized entangling operations could still be realized. The correct gate time for each \Wc  could be set by the time that the trapping wells are kept resonant, using DC control fields. Or, the reduced spacing $s_x = \SI{40}{\micro\meter}$ could be consecutively set for the different pairs of linear traps using multiple adjustments of the RF voltage. Another important improvement would be a further segmentation of the DC rails. This would allow for tuning of additional parameters such as the mode tilts and the axial trap spacing $s_z$.

\section{Advantages of RF tuning in linear trap arrays}
\label{app:advantages_RF_tuning}

For the successful application of a trap array for quantum information processing, the inter-site coupling rate \Wc needs to significantly exceed the motional heating rate \Gh. In surface traps, the heating rate typically increases drastically as the ions approach the trap surface \cite{Bro2015}. Therefore, one would like to maximize the ion-surface separation $d$ while maintaining a small trap spacing $s$ that yields a sufficient coupling rate \Wc, cf. equation~(\ref{eq:coupling-rate}). In this section, we consider a trap array without RF tuning and with a trap distance $s_x=\SI{40}{\micro\meter}$. For such an array, the ion-surface separation cannot exceed $d_\text{max}\approx\SI{30}{\micro\meter}$, as we show below. In contrast, using RF tuning we achieve a more than three times larger ion-surface separation $d\approx\SI{100}{\micro\meter}$ for the same ion-ion spacing $s_x$, cf. figure~\ref{fig_app:sim_big-array_shuttling}. This increased separation $d$ corresponds to a two orders of magnitude lower ion heating rate \Gh, assuming a typical $d^{-4}$ dependence of surface noise \cite{Bol2018,Sed2018-2}. In addition, the use of RF voltage tuning in linear trap arrays can also lead to a greatly increased trap depth, as shown below.
\begin{figure}[htbp]
	\centering
	\includegraphics[width=0.82\textwidth]{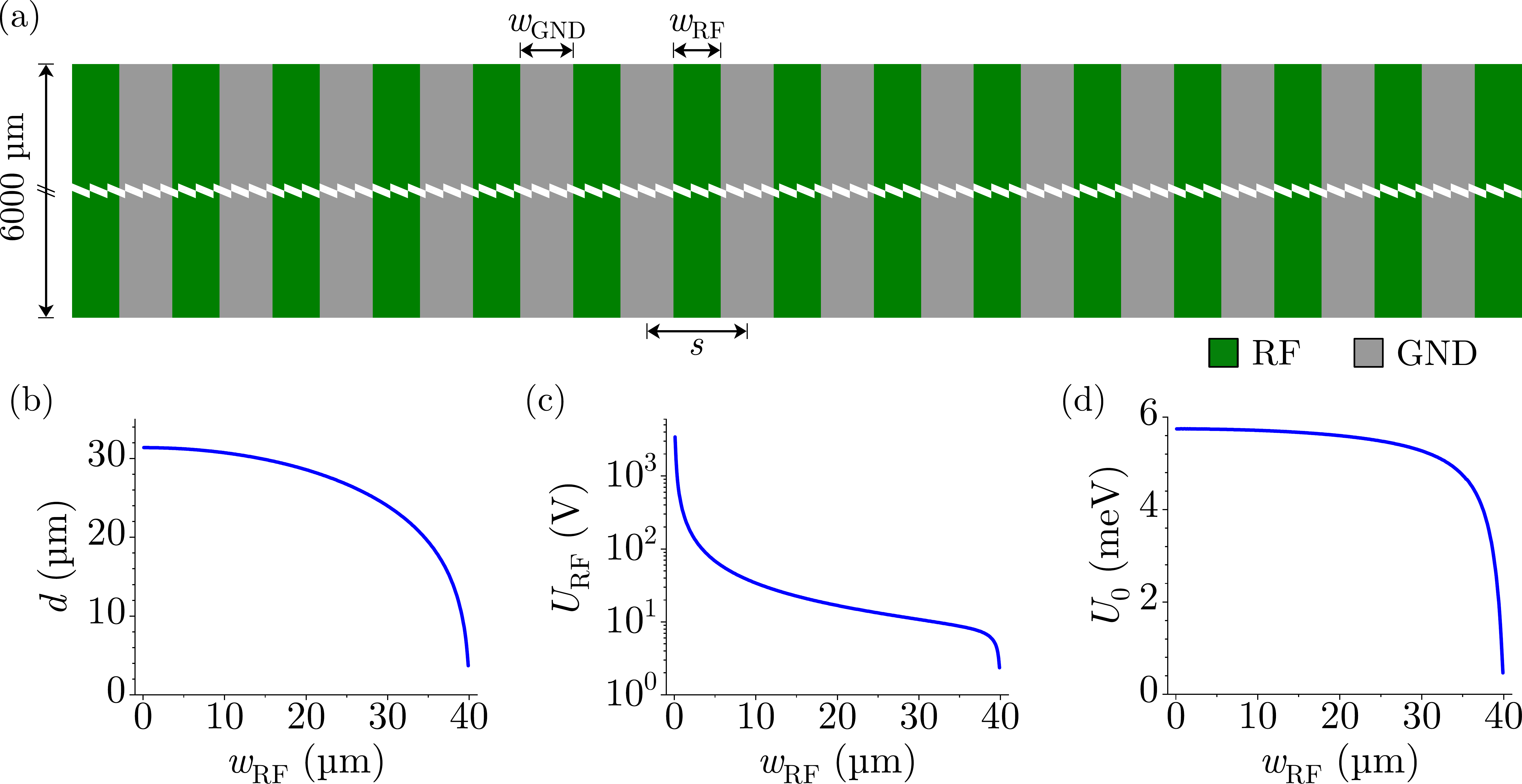}
	\caption{Linear trap array without RF tuning. (a) Electrode geometry with parallel RF and GND rails of widths $w_\text{RF}$ and $w_\text{GND}$, respectively. An additional grounded plane is at a distance $y=\SI{1}{\milli\meter}$ above the trap surface. (b) Ion-surface separation $d$ for a trapping site at the center of the array as function of the RF rail width $w_\text{RF}$. (c) RF voltage required to obtain a trap stability factor $q=0.4$, as function of $w_\text{RF}$. The RF drive frequency is $\Wrf=2\pi\times\SI{30}{\mega\hertz}$. (d) Corresponding trap depth $U_0$ for the same RF drive settings as in (c).}
	\label{fig_app:sim_static_array}	
\end{figure}

We consider a linear trap array with alternating RF and GND rails, figure~\ref{fig_app:sim_static_array}\,(a). An additional grounded plane at a distance $y=\SI{1}{\milli\meter}$ above the trap surface is assumed\footnote{The effect of the grounded plane above the trap surface on the trapping potentials is negligible due to the comparably small ion-surface separation $d$.}. The important difference to the array in figure~\ref{fig_app:geometry_big-array} is that the RF voltage is identical on all RF rails, i.e. there is no RF tuning. Therefore, the trap distance $s_x$ is simply given by the periodicity of the structure, $s_x=w_\text{GND}+w_\text{RF}$, neglecting edge effects. We further consider a fixed ion-ion spacing, $s_x = \SI{40}{\micro\meter}$, the same value as proposed for the ion-ion coupling in figure~\ref{fig_app:sim_big-array_shuttling}. We then simulate the trapping potential for different RF rail widths, spanning the entire range $w_\text{RF}\in(0,s_x)$. With $w_\text{GND}=s_x-w_\text{RF}$, the electrode geometry is thus fully determined. In this way, we find all possible values for the ion-surface separation $d$ that can be realized. As seen in (b), the maximum achievable ion-surface separation is about $d_\text{max}\approx\SI{30}{\micro\meter}$. This is more than a factor 3 smaller than the ion-surface separation $d\approx\SI{100}{\micro\meter}$ in appendix~\ref{app:large_array}, where RF shuttling is used to realize the ion-ion spacing $s_x = \SI{40}{\micro\meter}$. In the limit of thin RF rails, $w_\text{RF}\to0$, where $d_\text{max}$ is reached, the stable operation of the trap becomes increasingly inefficient. This is evidenced by the divergence of the RF drive voltage \Urf required for a fixed stability $q$-factor, figure~\ref{fig_app:sim_static_array}\,(c). We further determine the global trap depth $U_0$, shown in (d), which has a maximum value of $U_0\approx\SI{6}{\milli\electronvolt}$. While this is comparable to the double well barrier \Ub in figure~\ref{fig_app:sim_big-array_shuttling}, it is significantly smaller than the global depth $U_0$ of several hundred \si{\milli\electronvolt} in appendix~\ref{app:large_array}, making loading of ions extremely challenging.

\section{Suppression of parasitic motional coupling between non-nearest neighbors}
\label{app:parasitic_coupling}

For the realization of parallelized pairwise entangling operations between ions in adjacent trapping sites we suggest to employ the motional coupling of the ions’ axial modes. Using RF shuttling, a motional coupling strength $\Wc\approx2\pi\times\SI{1}{\kilo\hertz}$ between nearest neighbor ions can be reached at a reduced trap distance $s_x = \SI{40}{\micro\meter}$, see appendix~\ref{app:large_array}. However, the desired coherent evolution of the ions’ motional states can be disrupted by additional unwanted motional couplings to non-nearest neighbors. While these parasitic couplings become rapidly weaker for higher order neighbors due to the $1/s^3$ scaling of the motional coupling strength \Wc, cf. equation~(\ref{eq:coupling-rate}), their presence can still degrade the gate fidelities. In this section we outline a scheme to considerably reduce the parasitic coupling to non-nearest neighbor ions. The outline considers parallelized entangling operations along the $x$-direction; the scheme works in the same fashion for operations along the axial direction $z$.
\begin{figure}[htbp]
	\centering
	\includegraphics[width=\textwidth]{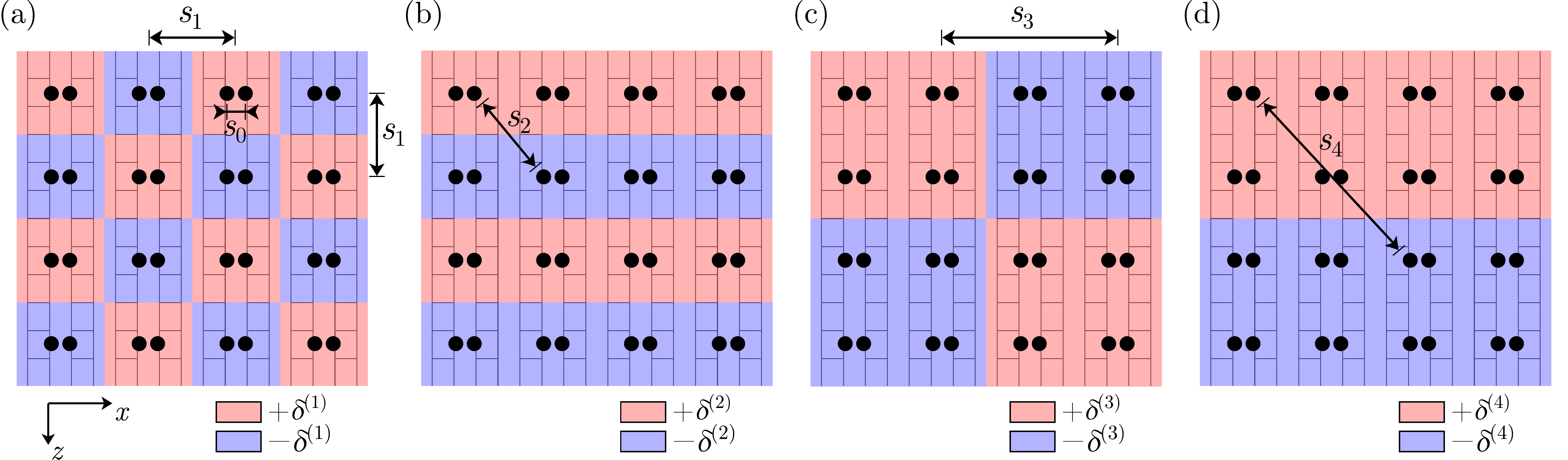}
	\caption{Suppression of parasitic motional coupling during parallelized entangling operations. The ion lattice is grouped in pairs of ions (black spheres) with distance $s_0$ and resonant confining double well potential. The trap electrodes are illustrated as gray lines. (a) First order parasitic coupling between ions at distance $s_1$ is suppressed by choosing alternating resonant double well frequencies $\wz+\delta^{(1)}$ (red squares) and $\wz-\delta^{(1)}$ (blue squares). (b) Second order parasitic coupling between ions at distance $s_2$ is suppressed by an additional detuning of rows of ions by $+\delta^{(2)}$ (red stripes) and $-\delta^{(2)}$ (blue stripes). For higher order couplings, additional detunings in successively increasing areas are required. The areas for the third and fourth order are depicted in (c) and (d), respectively. The detunings  $\delta^{(i)}$ decrease with the order of the coupling, scaling as $1/s_i^3$.}
	\label{fig_app:parasitic_coupling}	
\end{figure}

The scheme makes use of the fact that a strong coupling of the ions’ motion requires their axial well frequencies to be resonant; for a well detuning $\delta\gg\Wc$, the ion motion can be considered independent \cite{Bro2011,Har2011,Wil2014}. It is therefore possible to pairwise couple multiple ions simultaneously by picking a different resonance frequency for different ion pairs, effectively turning off the parasitic coupling between the non-nearest neighbors. To be more concrete, we consider the ion lattice illustrated in figure~\ref{fig_app:parasitic_coupling}. The ions are arranged in pairs with a nearest neighbor distance $s_0= \SI{40}{\micro\meter}$ along the $x$-direction, giving rise to a coupling strength $\Wc^{(0)}=2\pi\times\SI{1}{\kilo\hertz}$ at a resonant axial frequency $\wz\approx2\pi\times \SI{1}{\mega\hertz}$. The distance to the next-nearest neighbor ions is assumed to be $s_1=\SI{200}{\micro\meter}$, similar to the trapping potential in figure~\ref{fig_app:sim_big-array_shuttling}. For simplicity, we further assume an isotropic lattice, where the distance $s_1$ and the motional coupling strength \Wc are identical along the $x$- and $z$-direction\footnote{For the more general case of an anisotropic lattice, considered in appendix~\ref{app:large_array}, the scheme works in an analogue way. For a given order of parasitic coupling, the required detuning is given by the maximum of the couplings along the $x$- and $z$-direction.}. For the suppression of the first order parasitic coupling between ions at distance $s_1$, one can use a checkerboard pattern for the well detuning, as shown in figure~\ref{fig_app:parasitic_coupling}\,(a). Axial well frequencies within a red or blue square are detuned by $+\delta^{(1)}$ and $-\delta^{(1)}$, respectively, relative to the resonant well frequency \wz. For the required detuning it holds $\delta^{(1)}\gg\Wc^{(1)}=\Wc^{(0)} (s_0/s_1)^3=2\pi\times \SI{8}{\hertz}$. Already for $\delta^{(1)}=10\,\Wc^{(1)}\approx2\pi\times \SI{100}{\hertz}$, a drastic reduction of the parasitic coupling should be observable. The second order parasitic coupling is between ions at a distance $s_2=\sqrt{2}s_1$, along the diagonal of the lattice. This coupling can be suppressed with an additional detuning $\pm\delta^{(2)}$, applied to adjacent rows of ions on a striped pattern as shown in (b). Here,
$\delta^{(2)}\gg\Wc^{(2)}=\Wc^{(0)} (s_0/s_2)^3\approx2\pi\times \SI{2.8}{\hertz}$. It is important to note that the detuning $\delta^{(2)}$ partially cancels the detuning $\delta^{(1)}$ for some pairs of ions. To account for this cancellation, the detuning $\delta^{(1)}$ must be increased accordingly. Hence, the suppression of both first and second order parasitic couplings requires 4 different well frequencies for the pairs of ions across the array:  
$\wz+\delta^{(1)}+2\delta^{(2)}$, $\wz+\delta^{(1)}$, $\wz-\delta^{(1)}$, $\wz-\delta^{(1)}-2\delta^{(2)}$. The scheme can be further extended to suppress higher order couplings by successively increasing the cell size of the checkerboard and striped patterns, as shown in (c) and (d) for the third and fourth order coupling, respectively. We note that a complete suppression of parasitic couplings up to infinite order is impossible due to the partial cancellation of detunings for different orders: the accumulated compensation for the cancellation leads to diverging well frequencies in the limit of infinite order couplings. In practice, however, higher order couplings $i$ can be neglected once their coupling strength $\Wc^{(i)}$ falls below the ion heating rate \Gh, the fundamental limit for uncontrolled motional excitation. For instance, in the considered array already the fourth order coupling has a strength $\Wc^{(4)}<2\pi\times \SI{1}{\hertz}$. Suppressing parasitic coupling up to the fourth order requires 16 different well frequencies for adjacent pairs of ions with a maximum detuning from each other on the order of a few hundred Hz. Such detunings are small compared to the well frequency $\wz\approx2\pi\times \SI{1}{\mega\hertz}$ and can be readily implemented using individual DC control electrodes below each trapping site.



\bibliographystyle{MSP}
\bibliography{references}

\clearpage

\end{document}